\documentclass[letterpaper, 10 pt, conference]{ieeeconf}%
\usepackage{graphicx}
\usepackage{epsfig}
\usepackage{amsmath}
\usepackage{amssymb}
\usepackage{bbm}
\usepackage{float}
\usepackage{tikz}
\usepackage{algorithm}
\usepackage{algpseudocode}
\usepackage{pgfplots}
\usepackage{gnuplottex}
\usepackage{mathrsfs}
\usepackage{subfigure}
\usepackage{booktabs}
\usepackage{multirow}
\usepackage{amsfonts}
\usepackage{wrapfig}
\usepackage{epstopdf}
\usetikzlibrary{plotmarks}

\providecommand{\U}[1]{\protect\rule{.1in}{.1in}}
\providecommand{\U}[1]{\protect\rule{.1in}{.1in}}
\IEEEoverridecommandlockouts

\makeatletter
\let\@ORGmakecaption\@makecaption
\long
\def\@makecaption#1#2{\@ORGmakecaption{#1}{#2}\vskip\belowcaptionskip\relax}
\makeatother
\def\@normalsize{\@setsize\normalsize{10pt}\xpt\@xpt
\abovedisplayskip 10pt plus2pt minus5pt\belowdisplayskip
\abovedisplayskip \abovedisplayshortskip \z@
plus3pt\belowdisplayshortskip 6pt plus3pt
minus3pt\let\@listi\@listI}
\makeatletter
\newcommand\npar{\@startsection{section}{1pt}}
\makeatother

\newcommand{\bsym}[1]{\boldsymbol{#1}}
\newtheorem{theorem}{Theorem}
\newcounter{lemcount}

\newtheorem{corollary}[theorem]{Corollary}

\newcommand{\qed}{\nobreak \ifvmode \relax \else
\ifdim\lastskip<1.5em \hskip-\lastskip
\hskip1.5em plus0em minus0.5em \fi \nobreak
\vrule height0.4em width0.5em depth0.25em\fi}

\usetikzlibrary{decorations.pathmorphing,shadows}

\newcommand{\sgn}[1]{\mbox{sgn}(#1)}

\graphicspath{{C:/Users/Yas/Dropbox/Academic/Research/Fall2014/DH/Figures/}}

\title{{\LARGE \textbf{An Optimal Control Approach for the Data Harvesting Problem}}}
\author{ \parbox{3.5 in}{\centering Yasaman Khazaeni and Christos G. Cassandras\\
         Division of Systems Engineering\\ and Center for Information and Systems Engineering\\
         Boston University, MA 02446\\
         {\tt\small yas@bu.edu,cgc@bu.edu}}
         \thanks{The authors’ work is supported in part by NSF under grants CNS-
1239021, ECCS-1509084, and IIP-1430145, by AFOSR under grant FA9550-12-1-0113, by ONR under grant N00014-09-1-1051, and by the Cyprus Research Promotion
Foundation under Grant New Infrastructure Project/Strategic/0308/26.}
}
\allowdisplaybreaks
\begin{document}
\maketitle
\thispagestyle{empty}
\pagestyle{empty}
\begin{abstract}
We propose a new method for trajectory planning to solve the data harvesting problem. In a two-dimensional mission space, $N$ mobile agents are tasked with the collection of data generated at $M$ stationary sources and delivery to a base aiming at minimizing expected delays. An optimal control formulation of this problem provides some initial insights regarding its solution, but it is computationally intractable, especially in the case where the data generating processes are stochastic. We propose an agent trajectory parameterization in terms of general function families which can be subsequently optimized on line through the use of Infinitesimal Perturbation Analysis (IPA). Explicit results are provided for the case of elliptical and Fourier series trajectories and some properties of the solution are identified, including robustness with respect to the data generation processes and scalability in the size of an event set characterizing the underlying hybrid dynamic system.
\end{abstract}
\section{Introduction}
Systems consisting of cooperating mobile agents are being continuously
developed for a broad spectrum of applications such as environmental sampling
\cite{Corke2010},\cite{Smith2011}, surveillance \cite{Tang2005}, coverage
control \cite{Zhong2011},\cite{Chakrabarty2002},\cite{Cardei2005}, persistent
monitoring \cite{Alamdari2013},\cite{Cassandras2013_2}, task assignment
\cite{PanagouTK14}, and data harvesting and information collection
\cite{Klesh2008},\cite{Ny2008},\cite{Moazzez-Estanjini2012}. The data
harvesting problem arises in many settings, including \textquotedblleft smart
cities\textquotedblright\ where wireless sensor networks (WSNs) are being
widely deployed for purposes of monitoring the environment, traffic,
infrastructure for transportation and for energy distribution, surveillance,
and a variety of other specialized purposes \cite{Roscia2013}. Although many
efforts focus on the analysis of the vast amount of data gathered, we must first
ensure the existence of robust means to collect all data in a timely fashion
when the size of the sensor networks and the level of node interference do not allow for a fully wireless connected system. Sensors
can locally gather and buffer data, while mobile elements (e.g., vehicles,
aerial drones) retrieve the data from each part of the network. Similarly,
mobile elements may themselves be equipped with sensors and visit specific
points of interest to collect data which must then be delivered to a given
base. These mobile agents should follow an optimal path (in some sense to be
defined) which allows visiting each data source frequently enough and within
the constraints of a given environment like that of an urban setting.

The data harvesting problem using mobile agents known as \textquotedblleft
message ferries\textquotedblright\ or \textquotedblleft data
mules\textquotedblright\ has been considered from several different
perspectives. For a survey on different routing problems in WSNs see
\cite{Akkaya2005},\cite{Liu2011} and references therein. In \cite{Chang2014}
algorithms are proposed for patrolling target points with the goal of balanced
time intervals between consecutive visits. A weighted version of the algorithm
improves the performance in cases with unequally valued targets. However, in
this scenario the data need not be delivered to a base and visits to a
recharging station are only necessary if the data mules are running out of
energy. In \cite{Ny2008} the problem is viewed as a polling system with a
mobile server visiting data queues at fixed targets. Trajectories are designed
for the mobile server in order to stabilize the system, keeping queue contents
(modeled as fluid queues) uniformly bounded.

In this paper, we consider the data harvesting problem as an optimal control
problem for a team of multiple cooperating mobile agents responsible for
collecting data generated by arbitrary random processes\textbf{ }at fixed
target points and delivering these data to a base. The ultimate goal is for
the data to be collected and delivered with minimum expected delay. Rather
than looking at this problem as a scheduling task where visit times for each
target are determined assuming agents only move in straight lines between
targets, we aim to optimize a two-dimensional trajectory for each agent, which
may be periodic and can collect data from a target once the agent is within a
given range from that target. Interestingly, the setting of the problem can
also be viewed as an evacuation process where visits are needed to retrieve
individuals from a set of target points which may be of non-uniform
importance. In this paper, we limit ourselves to trajectories with no
constraints due to obstacles or other factors. Clearly, in an urban
environment this is generally not the case and the set of admissible
trajectories will have to be restricted in subsequent work.

We formulate a finite-horizon optimal control problem in which the underlying
dynamic system has hybrid (time-driven and event-driven) dynamics. We note
that the specification of an appropriate objective function is nontrivial for
the data harvesting problem, largely due to the fact that the agents act as
mobile servers for the data sources and have their own dynamics. Since the
control is applied to the motion of agents, the objective function must
capture the agent behavior in addition to that of the data queues at the
targets, the agents, and the base. The solution of this optimal control
problem (even in the deterministic case) requires a Two Point Boundary Value
Problem (TPBVP) numerical solver which is clearly not suited for on-line
operation and yields only locally optimal solutions. Thus, the main
contribution of the paper is to formulate and solve an optimal parametric
agent trajectory problem. In particular, similar to the idea in
\cite{Lin2014} we represent an agent trajectory in terms of general function
families characterized by a set of parameters that we seek to optimize, given
an objective function. We consider elliptical trajectories as well as the much
richer set of Fourier series trajectory representations. We then show that we
can make use of Infinitesimal Perturbation Analysis (IPA) for hybrid systems
\cite{Cassandras2010} to determine gradients of the objective function with
respect to these parameters and subsequently obtain (at least locally) optimal
trajectories. This approach also allows us to exploit $(i)$ robustness
properties of IPA to allow stochastic data generation processes, $(ii)$
the event-driven nature of the IPA gradient estimation process which is
scalable in the event set of the underlying hybrid dynamic system, and $(iii)$ the on-line computation which implies that trajectories adjust as operating conditions change (e.g., new targets).

In section \ref{Formulation} we present an optimal control formulation for the
data harvesting problem. In section \ref{OptimalControl} we provide a
Hamiltonian analysis leading to a TPBVP. In section
\ref{parametric} we formulate the alternative problem of determining optimal
trajectories based on general function representations and provide solutions
using a gradient-based algorithm using IPA for two particular function
families. Sections \ref{Numerical} and \ref{Conclusions} present the numerical
results and the conclusions respectively.

\section{Problem Formulation}

\label{Formulation} We consider a data harvesting problem where $N$ mobile
agents collect data from $M$ stationary targets in a two-dimensional
rectangular mission space $S=[0,L_{1}]\times\lbrack0,L_{2}]\subset
\mathbbm{R}^{2}$. Each agent may visit one or more of the $M$ targets, collect
data from them, and deliver them to a base. It then continues visiting
targets, possibly the same as before or new ones, and repeats this process. By
cooperating in how data are collected and delivered, the objective of the
agent team is to minimize a weighted sum of collection and delivery delays
over all targets.

Let $s_{j}(t)=[s_{j}^{x}(t),s_{j}^{y}(t)]$ be the position of agent $j$ at
time $t$ with $s_{j}^{x}(t)\in\lbrack0,L_{1}]$ and $s_{j}^{y}(t)\in
\lbrack0,L_{2}]$. The position of the agent follows single integrator
dynamics:
\begin{equation}
\dot{s}_{j}^{x}(t)=u_{j}(t)\cos\theta_{j}(t),\quad\ \ \dot{s}_{j}^{y}%
(t)=u_{j}(t)\sin\theta_{j}(t) \label{agentdynamics}%
\end{equation}
where $u_{j}(t)$ is the scalar speed of the agent (normalized so that $0\leq
u_{j}(t)\leq1$) and $\theta_{j}(t)$ is the angle relative to the positive
direction, $0\leq\theta_{j}(t)<2\pi$. Thus, we assume that the agent controls
its orientation and speed. An agent is represented as a particle, so that we
will omit the need for any collision avoidance control. The agent dynamics
above could be more complicated without affecting the essence of our analysis,
but we will limit ourselves here to (\ref{agentdynamics}).

Consider a set of data sources as points $w_{i}\in S,$ $i=1,\dots,M,$ with
associated ranges $r_{ij}$, so that agent $j$ can collect data from $w_{i}$
only if the Euclidean distance $D_{ij}(t)=\Vert w_{i}-s_{j}(t)\Vert$ satisfies
$D_{ij}(t)\leq r_{ij}$. Similarly, there is a base at $w_{\!_{B}}\in S$ which
receives all data collected by the agents. An agent can only deliver data to
the base if the Euclidean distance $D_{\!_{Bj}}(t)=\Vert w_{\!_{Bj}}%
-s_{j}(t)\Vert$ satisfies $D_{\!_{Bj}}(t)\leq r_{Bj}$. We define a function
$P_{ij}(t)$ to be the normalized data collection rate from target $i$ when the
agent is at $s_{j}(t)$:
\begin{equation}
P_{ij}(t)=p(w_{i},s_{j}(t))\label{Pij}%
\end{equation}
and we assume that: $({\bf A1})$ it is monotonically non-increasing in the value of $D_{ij}(t)=\Vert w_{i}-s_{j}(t)\Vert$, and $({\bf A2})$ it satisfies $P_{ij}(t)=0$
if $D_{ij}(t)>r_{ij}$. Thus, $P_{ij}(t)$ can model communication power
constraints which depend on the distance between a data source and an agent
equipped with a receiver (similar to the model used in \cite{Ny2008}) or
sensing range constraints if an agent collects data using on-board sensors.
For simplicity, we will also assume that: $({\bf A3})$ $P_{ij}(t)$ is continuous in
$D_{ij}(t)$. Similarly, we define:
\begin{equation}
P_{\!_{Bj}}(t)=p(w_{\!_{B}},s_{j}(t))\label{PB}%
\end{equation}
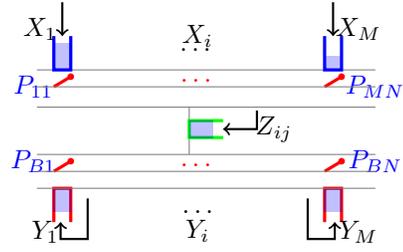
\begin{figure}[ptb]
\begin{center}
\begin{tikzpicture}[scale=0.45]
\draw[thin,gray](0,0)--(10,0);
\draw[very thick,blue] (0.5,1)--(0.5,0)--(1,0)--(1,1);
\draw [fill,blue, nearly transparent]  (0.5,0.8)--(0.5,0)--(1,0)--(1,0.8);
\draw [black,thick,->] (0.75,2.0)--(0.75,1.0);
\draw (0.1,1.2) node {$X_1$};
\draw (4.75,0.6) node {$\dots$};
\draw (4.75,1.0) node {$X_i$};
\draw[very thick,blue] (8.5,1)--(8.5,0)--(9,0)--(9,1);
\draw [fill,blue, nearly transparent]  (8.5,0.4)--(8.5,0)--(9,0)--(9,0.4);
\draw [black,thick,->] (8.75,2.0)--(8.75,1.0);
\draw (9.5,1.2) node {$X_M$};
\draw[blue] (-0.1,-0.5) node {$P_{11}$};
\draw[blue] (10.0,-0.5) node {$P_{MN}$};
\draw[very thick,red] (0.5,-0.5)--(1,-0.2);
\draw[very thick,red,fill] (1,-0.2) circle(0.05);
\draw[very thick,red] (4.7,-0.3) node {$\ldots$};
\draw[very thick,red] (8.5,-0.5)--(9,-0.2);
\draw[very thick,red,fill] (9,-0.2) circle(0.05);
\draw[thin,gray](0,-1.1)--(10,-1.1);
\draw[thin,gray](0,-0.5)--(10,-0.5);
\draw[thin,gray](4.5,-1.1)--(4.5,-2.5);
\draw[very thick,green] (5.5,-1.5)--(4.5,-1.5)--(4.5,-2)--(5.5,-2);
\draw [fill,blue, nearly transparent]  (5.2,-1.5)--(4.5,-1.5)--(4.5,-2)--(5.2,-2);
\draw [black,thick,->] (6.5,-1.2)--(6.5,-1.75)--(5.5,-1.75);
\draw (7,-1.75) node {$Z_{ij}$};
\draw[thin,gray](0,-2.5)--(10,-2.5);
\draw[blue] (-0.1,-2.7) node {$P_{B1}$};
\draw[blue] (10.0,-2.7) node {$P_{BN}$};
\draw[very thick,red] (0.5,-3)--(1,-2.7);
\draw[very thick,red,fill] (1,-2.7) circle(0.05);
\draw[very thick,red] (4.7,-2.8) node {$\ldots$};
\draw[very thick,red] (8.5,-3.0)--(9,-2.7);
\draw[very thick,red,fill] (9,-2.7) circle(0.05);
\draw[thin,gray](0,-3)--(10,-3);
\draw[very thick,red] (0.5,-4.5)--(0.5,-3.5)--(1,-3.5)--(1,-4.5);
\draw [fill,blue, nearly transparent]  (0.5,-4.2)--(0.5,-3.5)--(1,-3.5)--(1,-4.2);
\draw [black,thick,->] (1.5,-3.8)--(1.5,-5)--(0.7,-5)--(0.7,-4.5);
\draw (0.25,-4.8) node {$Y_{1}$};
\draw (4.75,-4.2) node {$\ldots$};
\draw[very thick,red] (8.5,-4.5)--(8.5,-3.5)--(9,-3.5)--(9,-4.5);
\draw [fill,blue, nearly transparent]  (8.5,-4.2)--(8.5,-3.5)--(9,-3.5)--(9,-4.2);
\draw [black,thick,->] (8,-3.8)--(8,-5)--(8.8,-5)--(8.8,-4.5);
\draw (9.5,-4.8) node {$Y_{M}$};
\draw[thin,gray](0,-3.5)--(10,-3.5);
\draw (4.75,-4.75) node {$Y_{i}$};
\end{tikzpicture}
\end{center}
\caption{Data harvesting queueing model for $M$ targets and $N$ agents}%
\label{queueschematic}%
\end{figure}The data harvesting problem described above can be viewed as a
polling system where mobile agents are serving the targets by collecting data
and delivering them to the base. Figure \ref{queueschematic} shows a queueing
system in which each $P_{ij}(t)$ is depicted as a switch activated when
$D_{ij}(t)\leq r_{ij}$ to capture the finite range between agent $j$ and
target $i$. All queues are modeled as flow systems whose dynamics are given
next (however, as we will see, the agent trajectory optimization is driven by
events observed in the underlying system where queues contain discrete data
packets so that this modeling device has minimal effect on our analysis). As
seen in Fig. \ref{queueschematic}, there are three sets of queues. The first
set includes the data contents $X_{i}(t)\in\mathbb{R}^{+}$ at each target
$i=1,...,M$ where we use $\sigma_{i}(t)$ as the instantaneous inflow rate. In
general, we treat $\{\sigma_{i}(t)\}$ as a random process assumed only to be
piecewise continuous; we will treat it as a deterministic constant only for
the Hamiltonian analysis in the next section. Thus, at time $t$, $X_{i}(t)$ is
a random variable resulting from the random process $\{\sigma_{i}(t)\}$. The
second set of queues consists of data contents $Z_{ij}(t)\in\mathbb{R}^{+}$
onboard agent $j$ collected from target $i$ as long as $P_{ij}(t)>0$. The
last set consists of queues $Y_{i}(t)\in\mathbb{R}^{+}$ containing data at the
base, one queue for each target, delivered by some agent $j$ as
long as $P_{\!_{Bj}}(t)>0$. Note that $\{X_i(t)\}$, $\{Z_{ij}(t)\}$ and $\{Y_{i}(t)\}$
are also random processes and the same applies to the agent states
$\{s_{j}(t)\}$, $j=1,\dots,N$, since the controls are generally dependent on
the random queue states. Thus, we ensure that all random processes are defined
on a common probability space. The maximum rate of data collection from target
$i$ by agent $j$ is $\mu_{ij}$ and the actual rate is $\mu_{ij}P_{ij}(t)$ if
$j$ is connected to $i$. We will assume that: $({\bf A4})$ only one agent at a time is
connected to a target $i$ even if there are other agents $l$ with
$P_{il}(t)>0$; this is not the only possible model, but we adopt it based on
the premise that simultaneous downloading of packets from a common source
creates problems of proper data reconstruction at the base. The dynamics of
$X_{i}(t)$, assuming that agent $j$ is connected to it, are
\begin{equation}
\dot{X}_{i}(t)=\left\{
\begin{array}
[c]{ll}%
0 \qquad\quad \mbox{if }X_{i}(t)=0\mbox{ and }\sigma_{i}(t)\leq\mu_{ij}(t)P_{ij}(t)\\
\sigma_{i}(t)-\mu_{ij}(t)P_{ij}(t)  \quad\qquad\qquad\qquad \mbox{otherwise}
\end{array}
\right.  \label{Xdot}%
\end{equation}
Obviously, $\dot{X}_{i}(t)=\sigma_{i}(t)$ if $P_{ij}(t)=0$, $j=1,\dots,N$. In order to express the dynamics of $Z_{ij}(t)$, let
\begin{equation}
\resizebox{0.99 \columnwidth}{!}{$ \tilde\mu_{ij}(t)=\left\{\begin{array}{l l} \min\Big(\frac{\sigma_i(t}{P_{ij}(t)},\mu_{ij}\Big) \quad &\mbox{if }X_{i}(t)=0 \text{ and } P_{ij}(t)>0\\ \mu_{ij} \quad &\mbox{otherwise}\end{array} \right. $}\label{mutilde}%
\end{equation}
This gives us the dynamics:
\begin{equation}
\resizebox{1.0 \columnwidth}{!}{$ \dot Z_{ij}(t)=\left\{ \begin{array}{l l} 0 \qquad \mbox{if }Z_{ij}(t)=0 \mbox{ and } \tilde\mu_{ij}(t)P_{ij}(t)-\beta_{ij}P_{\!_{Bj}}(t)\le 0 \\ \tilde\mu_{ij}(t)P_{ij}(t)-\beta_{ij}P_{\!_{Bj}}(t)  \qquad \qquad\qquad\qquad\mbox{otherwise} \end{array} \right.$}\label{Zdot}%
\end{equation}
where $\beta_{ij}$ is the maximum rate of data from target $i$ delivered by
agent $j$. For simplicity, we assume that: $({\bf A5})$ $\Vert w_{i}-w_{B}\Vert
>r_{ij}+r_{Bj}$ for all $i=1,\dots,M$ and $j=1,\dots,N$, i.e., the agent cannot collect and deliver data at the same time. Therefore, in (\ref{Zdot}) it is always the
case that $P_{ij}(t)P_{Bj}(t)=0$. Finally, the dynamics of $Y_{i}(t)$ depend
on $Z_{ij}(t)$, the content of the on-board queue of each agent $j$ from
target $i$ as long as $P_{Bj}(t)>0$. We define $\beta_{i}(t)=\sum_{j=1}%
^{N}\beta_{ij}P_{Bj}(t)\mathbf{1}[Z_{ij}(t)>0]$ to be the total instantaneous
delivery rate for target $i$ data, so that the dynamics of $Y_{i}(t)$ are:
\begin{equation}
\dot{Y}_{i}(t)=\beta_{i}(t)\label{Ydot}%
\end{equation}
Our objective is to maintain minimal values for all target and on-board agent
data queues, while maximizing the contents of the delivered data at the base
queues. Thus, we define $J_{1}(X_{1},\ldots,X_{M},t)$ to be the weighted sum
of expected target queue contents (recalling that $\{\sigma_i(t)\}$ are random processes):
\begin{equation}
J_{1}(X_{1},\ldots X_{M},t)=\sum\limits_{i=1}^{M}q_{i}E[X_{i}(t)]\label{J1}%
\end{equation}
where the weight $q_{i}$ represents the importance factor of target $i$. Similarly, we define a
weighted sum of expected base queues contents:
\begin{equation}
J_{2}(Y_{1},\ldots Y_{M},t)=\sum\limits_{i=1}^{M}q_{i}E[Y_{i}(t)]\label{J2}%
\end{equation}
For simplicity, we will in the sequel assume that $q_{i}=1$ for all $i$
without affecting any aspect of our analysis. Therefore, our optimization
objective may be a convex combination of (\ref{J1}) and (\ref{J2}). In
addition, we need to ensure that the agents are controlled so as to maximize
their utilization, i.e., the fraction of time spent performing a useful task
by being within range of a target or the base. Equivalently, we aim to
minimize the non-productive idling time of each agent during which it is not
visiting any target or the base. Let
\begin{equation}
\resizebox{1.0 \columnwidth}{!}{$
D_{ij}^{+}(t)=\max(0,D_{ij}(t)-r_{ij}),\text{\ }D_{\!_{Bj}}^{+}(t)=\max
(0,D_{\!_{Bj}}(t)-r_{\!_{Bj}})$}\label{Dplus}
\end{equation}
so that the idling time for agent $j$ occurs when $D_{ij}^{+}(t)>0$ for all
$i$ and $D_{Bj}^{+}(t)>0$. We define the idling function $I_{j}(t)$:
\begin{equation}
I_{j}(t)=\log\Bigg(1+D_{\!_{Bj}}^{+}(t)\prod_{i=1}^{M}D_{ij}^{+}%
(t)\Bigg)\label{Ij}%
\end{equation}
This function has the following properties. First, $I_{j}(t)=0$ if and only if
the product term inside the bracket is zero, i.e., agent $j$ is visiting a
target or the base; otherwise, $I_{j}(t)>0$. Second, $I_{j}(t)$ is
monotonically nondecreasing in the number of targets $M$. The logarithmic
function is selected so as to prevent the value of $I_{j}(t)$ from dominating
those of $J_{1}(\cdot)$ and $J_{2}(\cdot)$ when included in a single
objective function. We define:
\begin{equation}
J_{3}(t)=M_{I}\sum\limits_{j=1}^{N}E[I_{j}(t)]\label{J3}%
\end{equation}
where $M_{I}$ is a weight for the idling time effect relative to $J_{1}%
(\cdot)$ and $J_{2}(\cdot)$. Note that $I_{j}(t)$ is also a random variable
since it is a function of the agent states $s_{j}(t)$, $j=1,\dots,N$. Finally,
we define a terminal cost at $T$ capturing the expected value of the amount of
data left on board the agents, noting that the effect of this term vanishes as
$T$ goes to infinity as long as all $E[Z_{ij}(T)]$ remain bounded:
\begin{equation}
J_{f}(T)=\frac{1}{T}\sum\limits_{i=1}^{M}\sum\limits_{j=1}^{N}E[Z_{ij}%
(T)]\label{Jf}%
\end{equation}
We can now formulate a stochastic optimization problem $\mathbf{P1}$ where the
control variables are the agent speeds and headings denoted by the vectors
$\mathbf{u}(t)=[u_{1}(t),\dots,u_{N}(t)]$ and $\boldsymbol{\theta}%
(t)=[\theta_{1}(t),\dots,\theta_{N}(t)]$ respectively (omitting their
dependence on the full system state at $t$). We combine the objective function
components in \eqref{J1}, \eqref{J2}, \eqref{J3} and \eqref{Jf} to obtain:
\begin{equation}
\resizebox{1.0 \columnwidth}{!}{$ \min\limits_{\bf u(t),\bsym\theta(t)} J(T)=\frac{1}{T}\int_0^T\Big(\alpha J_1(t)-(1-\alpha)J_2(t)+J_3(t)\Big)+J_f(T)$}\label{GenOptim}%
\end{equation}
where $\alpha\in\lbrack0,1]$ is a weight capturing the relative
importance of collected data as opposed to delivered data and $0\leq
u_{j}(t)\leq1$, $0\leq\theta_{j}(t)<2\pi$. To simplify notation, we have also
expressed $J_{1}(X_{1},\ldots X_{M},t)$ and $J_{2}(Y_{1},\ldots Y_{M},t)$ as
$J_{1}(t)$ and $J_{2}(t)$.

Since we are considering a finite time optimization problem, instability in
the queues is not an issue. However, stability of such a system can indeed be
an issue in the sense of guaranteeing that $E[X_{i}(t)]<\infty$, $E[Z_{ij}(T)]<\infty$ for all $i,j$ under a particular
control policy when $t\to \infty$. This problem is considered in \cite{Ny2008} for a simpler
deterministic data harvesting model where target queues are required to be
bounded. In this paper, we do not explicitly study this issue; however, given a certain number of agents, it is possible to stabilize a
target queue by designing agent trajectories to ensure that the queue is
visited frequently enough and periodically emptied.
\section{Optimal Control Solution}
\label{OptimalControl} In this section, we address $\mathbf{P1}$ in a
setting where all data arrival processes are deterministic, so that all expectations in \eqref{J1}-\eqref{Jf} degenerate to their
arguments. We proceed with a standard Hamiltonian analysis leading to a Two
Point Boundary Value Problem (TPBVP) \cite{bryson1975applied} where the states
and costates are known at $t=0$ and $t=T$ respectively. We define a state
vector and the associated costate vector:
\[%
\begin{split}
\mathbf{X}&(t)=[    X_{1}(t),\dots,X_{M}(t),Y_{1}(t),\dots,Y_{M}(t),\\
&  Z_{11}(t),\dots,Z_{MN}(t),s_{1}^{x}(t),s_{1}^{y}(t),\dots,s_{N}%
^{x}(t),s_{N}^{y}(t)]
\end{split}
\]%
\[%
\begin{split}
\boldsymbol{\lambda}&(t)=[  \lambda_{1}(t),\dots,\lambda_{M}(t),\gamma
_{1}(t),\dots,\gamma_{M}(t),\\
&  \phi_{11}(t),\dots,\phi_{MN}(t),\eta_{1}^{x}(t),\eta_{1}^{y}(t),\dots
,\eta_{N}^{x}(t),\eta_{N}^{y}(t)]
\end{split}
\]
The Hamiltonian is
\begin{equation}
\resizebox{0.95 \columnwidth}{!}{$ \begin{split} &H({\bf X},{\bsym \lambda}, {\bf u},{\bsym \theta})=\frac{1}{T}\Big[\alpha J_1(t)-(1-\alpha)J_2(t)+J_3(t)\Big]\\ &+\sum_i\lambda_i(t)\dot X_i(t)+\sum_i\gamma_{i}(t)\dot Y_{i}(t)+\sum_i\sum_j\phi_{ij}(t)\dot Z_{ij}(t)\\ &+\sum_j\big(\eta_j^x(t) u_j(t)\cos\theta_j(t)+\eta_j^y(t) u_j(t)\sin\theta_j(t)\big) \end{split}$}
\label{hamiltonian}%
\end{equation}
where the costate equations are
\[%
\begin{array}
[c]{ll}%
\dot{\lambda}_{i}(t)=-\frac{\partial H}{\partial{X_{i}}}=-\frac{\alpha}{T} &
\lambda_{i}(T)=0\\
\dot{\gamma}_{i}(t)=-\frac{\partial H}{\partial{Y_{i}}}=\frac{1-\alpha}{T} &
\gamma_{i}(T)=0\\
\dot{\phi}_{ij}(t)=-\frac{\partial H}{\partial{Z_{ij}}}=0 & \phi_{ij}%
(T)=\frac{\partial J_{f}(t)}{\partial Z_{ij}}\Big|_{T}%
\end{array}
\]%
\[
\resizebox{0.9 \columnwidth}{!}{$ \begin{split} \dot\eta_j^x(t)=&-\frac{\partial H}{\partial s_j^x}=-\Bigg[\frac{M_I}{T}\frac{\partial I_j(t)}{\partial s_j^x}+\sum_i\frac{\partial }{\partial s_j^x}\lambda_i(t)\dot X_i(t)\\ &+\sum_i\frac{\partial}{\partial s_j^x} \gamma_{i}(t)\dot Y_{i}(t)+\sum_i\frac{\partial}{\partial s_j^x} \phi_{ij}(t)\dot Z_{ij}(t)\Bigg] \end{split} $}
\]%
\begin{equation}
\resizebox{0.9 \columnwidth}{!}{$ \begin{split} \dot\eta_j^y(t)=&-\frac{\partial H}{\partial s_j^y}=-\Bigg[\frac{M_I}{T}\frac{\partial I_j(t)}{\partial s_j^y}+\sum_i\frac{\partial }{\partial s_j^y}\lambda_i(t)\dot X_i(t)\\ &+\sum_i\frac{\partial}{\partial s_j^y} \gamma_{i}(t)\dot Y_{i}(t)+\sum_i\frac{\partial}{\partial s_j^y} \phi_{ij}(t)\dot Z_{ij}(t)\Bigg] \end{split} $}
\label{costate}%
\end{equation}%
\[
\eta_{j}^{x}(T)=\eta_{j}^{y}(T)=0
\]
From \eqref{hamiltonian}, after some trigonometric manipulations, we get
\begin{equation}
\resizebox{1.0\hsize}{!}{$ \begin{split}& H({\bf X},{\bsym \lambda}, {\bf u},{\bsym \theta})=\frac{1}{T}\Big[\alpha J_1(t)-(1-\alpha)J_2(t)+J_3(t)\Big]\\&+\sum_i\lambda_i(t)\dot X_i(t)+\sum_i\gamma_{i}(t)\dot Y_{i}(t)+\sum_i\sum_j\phi_{ij}(t)\dot Z_{ij}(t)\\&+\sum_j u_j(t)\sgn{\eta_j^y(t)}\sqrt{{\eta_j^x(t)}^2+{\eta_j^y(t)}^2}\sin(\theta_j(t)+\psi_j(t)) \end{split}$}
\label{HamiltonianU}%
\end{equation}
where $\tan\psi_{j}(t)=\frac{\eta_{j}^{x}(t)}{\eta_{j}^{y}(t)}$ for $\eta
_{j}^{y}(t)\neq0$ and $\psi_{j}(t)=\sgn{\eta_j^x(t)}\frac{\pi}{2}$ if
$\eta_{j}^{y}(t)=0$. Applying the Pontryagin principle to \eqref{hamiltonian}
with $(\mathbf{u}^{\ast},{\boldsymbol{\theta}}^{\ast})$ being the optimal
control, we have:
\begin{equation}
H(\mathbf{X}^{\ast},{\boldsymbol{\lambda}}^{\ast},\mathbf{u}^{\ast
},{\boldsymbol{\theta}}^{\ast})=\min\limits_{\mathbf{u}(t),\boldsymbol{\theta
}(t)}H(\mathbf{X},{\boldsymbol{\lambda}},\mathbf{u},{\boldsymbol{\theta}})
\end{equation}
From \eqref{HamiltonianU} we easily see that we can always make the $u_{j}(t)$
multiplier to be negative, hence, recalling that $0\leq u_{j}(t)\leq1$,
\begin{equation}
u_{j}^{\ast}(t)=1 \label{OptimU}%
\end{equation}
Following the Hamiltonian definition in \eqref{hamiltonian} we have:
\begin{equation}
\frac{\partial H}{\partial\theta_{j}}=-\eta_{j}^{x}(t)u_{j}(t)\sin\theta
_{j}(t)+\eta_{j}^{y}(t)u_{j}(t)\cos\theta_{j}(t) \label{dHdtheta}%
\end{equation}
and setting $\frac{\partial H}{\partial\theta_{j}}=0$ the optimal heading
$\theta_{j}^{\ast}(t)$ should satisfy:

\begin{equation}
\tan\theta_{j}^{\ast}(t)=\frac{{\eta_{j}^{y}}(t)}{{\eta_{j}^{x}}(t)}%
\end{equation}
Since $u_{j}^{\ast}(t)=1$, we only need to evaluate $\theta_{j}^{\ast}(t)$ for
all $t\in\lbrack0,T]$. This is accomplished by discretizing the problem in
time and numerically solving a TPBVP with a forward integration of the state and a
backward integration of the costate. Solving this problem quickly becomes intractable as the number of agents and
targets grows. However, one of the insights this analysis provides is that
under optimal control the data harvesting process operates as a hybrid system
with discrete states (modes) defined by the dynamics of the flow
queues in \eqref{Xdot}, \eqref{Zdot}, \eqref{Ydot}, while the agents
maintain a fixed speed. The events that trigger mode transitions are
defined in Table \ref{eventlist} (the superscript $0$ denotes events
causing a variable to reach a value of zero from above and the superscript $+$
denotes events causing a variable to become strictly positive from a zero
value):\vspace{-4mm} \begin{table}[!htb]
\caption{Hybrid System Events}%
\label{eventlist}%
\centering
\begin{tabular}
[c]{|c|l|}\hline
Event Name & Description\\\hline
1. $\xi_{i}^{0}$ & $X_{i}(t)$ hits 0, for $i=1,\dots,M$\\\hline
2. $\xi_{i}^{+}$ & $X_{i}(t)$ leaves 0, for $i=1,\ldots,M$.\\\hline
3. $\zeta_{ij}^{0}$ & $Z_{ij}(t)$ hits 0, for $i=1,\ldots,M$, $j=1,\ldots
,N$\\\hline
4. $\delta_{ij}^{+}$ & $D_{ij}^{+}(t)$ leaves 0, for $i=1,\ldots,M$,
$j=1,\ldots,N$\\\hline
5. $\delta_{ij}^{0}$ & $D_{ij}^{+}(t)$ hits 0, for $i=1,\ldots,M$,
$j=1,\ldots,N$\\\hline
6. $\Delta_{j}^{+}$ & $D_{\!_{Bj}}^{+}(t)$ leaves 0, for $j=1,\ldots
,N$\\\hline
7. $\Delta_{j}^{0}$ & $D_{\!_{Bj}}^{+}(t)$ hits 0, for $j=1,\ldots,N$\\\hline
\end{tabular}
\end{table}

Observe that each of these events causes a change in at least one of the state
dynamics in \eqref{Xdot}, \eqref{Zdot}, \eqref{Ydot}. For example, $\xi
_{i}^{0}$ causes a switch in (\ref{Xdot}) from $\dot{X}_{i}(t)=\sigma
_{i}(t)-\mu_{ij}P_{ij}(t)$ to $\dot{X}_{i}(t)=0$. Also note that we have omitted an event $\zeta_{ij}^{+}$ for $Z_{ij}(t)$ leaving 0 since this event is
immediately induced by $\delta_{ij}^{0}$ when agent $j$ comes within range of
target $i$ and starts collecting data causing $Z_{ij}(t)$ to become positive
if $Z_{ij}(t)=0$ and $X_{i}(t)>0$. Finally, note that all events above are
directly observable during the execution of any agent trajectory and they do
not depend on our model of flow queues. For example, if $X_{i}(t)$ becomes
zero, this defines event $\xi_{i}^{0}$ regardless of whether the corresponding
queue is based on a flow or on discrete data packets; this observation is very
useful in the sequel.

The fact that we are dealing with a hybrid dynamic system further complicates
the solution of a TPBVP. On the other hand, it enables us to make use of
Infinitesimal Perturbation Analysis (IPA) \cite{Cassandras2010} to carry out
the parametric trajectory optimization process discussed in the next section.
In particular, we propose a parameterization of agent trajectories allowing us
to utilize IPA to obtain a gradient of the objective function with respect to
the trajectory parameters.

\section{Agent Trajectory Parameterization and Optimization}

\label{parametric} The idea here is to represent each agent's trajectory
through general parametric equations
\begin{equation}%
\begin{array}
[c]{ll}%
s_{j}^{x}(t)=f(\Theta_{j},\rho_{j}(t)),\text{ \ }\quad s_{j}^{y}%
(t)=g(\Theta_{j},\rho_{j}(t)) &
\end{array}
\label{param_traj}%
\end{equation}
where the function $\rho_j(t)$ controls the position of the agent on its
trajectory at time $t$ and $\Theta_{j}$ is a vector of parameters
controlling the shape and location of the agent $j$ trajectory. Let
$\Theta=[\Theta_{1},\dots,\Theta_{N}]$. We now replace problem
$\mathbf{P1}$ in (\ref{GenOptim}) by problem $\mathbf{P2}$:
\begin{equation}%
\begin{split}
\min\limits_{\Theta\in F_{\Theta}}  &  \frac{1}{T}\int_{0}^{T}\Big[\alpha J_{1}%
(\Theta,t)-(1-\alpha)J_{2}(\Theta,t)+J_{3}(\Theta,t)\Big]dt\\
&  +J_{f}(\Theta,T)
\end{split}
\label{ParamOptim}%
\end{equation}
where we return to allowing arbitrary stochastic data arrival processes
$\{\sigma_{i}(t)\}$ so that $\mathbf{P2}$ is a parametric stochastic
optimization problem with $F_{\Theta}$ appropriately defined depending on \eqref{param_traj}. The cost function in (\ref{ParamOptim}) is written as%
\[
J(\Theta,T;\mathbf{X}(\Theta,0))=E[\mathcal{L}(\Theta,T;\mathbf{X}%
(\Theta,0))]
\]
where $\mathcal{L}(\Theta,T;\mathbf{X}(\Theta,0))$ is a sample function
defined over $[0,T]$ and $\mathbf{X}(\Theta,0)$ is the initial value of the
state vector. For convenience, in the sequel we will use $\mathcal{L}_{1}$,
$\mathcal{L}_{2}$, $\mathcal{L}_{3}$, $\mathcal{L}_{f}$ to denote sample
functions of $J_{1}$, $J_{2}$, $J_{3}$ and $J_{f}$ respectively. Note that in
(\ref{ParamOptim}) we suppress the dependence of the four objective function
components on the controls $\mathbf{u}(t)$ and $\boldsymbol{\theta}(t)$ and
stress instead their dependence on the parameter vector $\Theta$. In the
rest of the paper, we will consider two families of trajectories motivated by
a similar approach used in the multi-agent persistent monitoring problem in
\cite{Lin2015}: \emph{elliptical} trajectories and a \emph{Fourier series}
trajectory representation which is more general and better suited for
non-uniform target topologies. The hybrid dynamics of the data harvesting
system allow us to apply the theory of IPA \cite{Cassandras2010} to obtain on
line the gradient of the sample function $\mathcal{L}(\Theta,T;\mathbf{X}%
(\Theta,0))$ with respect to $\Theta$. The value of the IPA\ approach
is twofold: $(i)$ The sample gradient $\nabla\mathcal{L}(\Theta,T)$ can be obtained on line based on observable sample path
data \emph{only}, and $(ii)$ $\nabla\mathcal{L}(\Theta,T)$ is an unbiased
estimate of $\nabla J(\Theta,T)$ under mild technical conditions as shown in
\cite{Cassandras2010}. Therefore, we can use $\nabla\mathcal{L}(\Theta,T)$
in a standard gradient-based stochastic optimization algorithm
\begin{equation}
\Theta^{l+1}=\Theta^{l}-\boldsymbol{\nu}_{l}\nabla\mathcal{L}(\Theta
^{l},T),\text{ \ }l=0,1,\ldots\label{SAalgo}%
\end{equation}
to converge (at least locally) to an optimal parameter vector $\Theta^{\ast
}$ with a proper selection of a step-size sequence $\{\boldsymbol{\nu}_{l}\}$ \cite{Kushner2003}. We emphasize that this process is carried out \emph{on
line}, i.e., the gradient is evaluated by observing a trajectory with given
$\Theta$ over $[0,T]$ and is iteratively adjusting it until
convergence is attained.

\subsubsection{IPA equations}

Based on the events defined earlier, we will specify event time derivative and
state derivative dynamics for each mode of the hybrid system. In this process,
we will use the IPA notation from \cite{Cassandras2010} so that ${\tau_{k}}$ is
the $k$th event time in an observed sample path of the hybrid system and
${\tau_{k}^{^{\prime}}}=\frac{d\tau_{k}}{d\Theta}$, $\mathcal{X}^{\prime
}(t)=\frac{d\mathcal{X}}{d\Theta}$ are the Jacobian matrices of partial
derivatives with respect to all components of the controllable parameter
vector $\Theta$. Throughout the analysis we will be using $(\cdot)^{\prime}$
to show such derivatives. We will also use $f_{k}(t)=\frac{d\mathcal{X}}{dt}$
to denote the state dynamics in effect over an interevent time interval
$[{\tau_{k},\tau_{k+1})}$. We review next the three fundamental IPA equations
from \cite{Cassandras2010} based on which we will proceed. First, events may
be classified as exogenous or endogenous. An event is exogenous if its
occurrence time is independent of the parameter $\Theta$, hence ${\tau
_{k}^{^{\prime}}}=0$. Otherwise, an endogenous event takes place when a
condition $g_{k}(\Theta,\mathcal{X})=0$ is satisfied, i.e., the state
$\mathcal{X}(t)$ reaches a switching surface described by $g_{k}%
(\Theta,\mathcal{X})$. In this case, it is shown in \cite{Cassandras2010} that
\begin{equation}
{\tau_{k}^{^{\prime}}}=-\Big(\frac{dg_{k}}{d\mathcal{X}}f_{k}(\tau_{k}%
^{-})\Big)^{-1}\Big(\frac{dg_{k}}{d\Theta}+\frac{dg_{k}}{d\mathcal{X}%
}{\mathcal{X}^{\prime}(\tau_{k}^{-})}\Big) \label{tauprime}%
\end{equation}
as long as $\frac{\partial g_{k}}{\partial\mathcal{X}}f_{k}(\tau_{k}^{-}%
)\neq0$. It is also shown in \cite{Cassandras2010} that the state derivative
$\mathcal{X}^{\prime}(t)$ satisfies
\begin{equation}
\frac{d}{dt}\mathcal{X}^{\prime}(t)=\frac{df_{k}}{d\mathcal{X}}\mathcal{X}%
^{\prime}(t)+\frac{d{f_{k}}}{d\Theta},\text{ \ \ }t\in\lbrack{\tau_{k}%
,\tau_{k+1})} \label{dXdtprime}%
\end{equation}
\begin{equation}
\mathcal{X}^{\prime}(\tau_{k}^{+})=\mathcal{X}^{\prime}(\tau_{k}^{-}%
)+[f_{k-1}(\tau_{k}^{-})-f_{k}(\tau_{k}^{+})]{\tau_{k}}^{\prime}
\label{Xprime}%
\end{equation}
Then, $\mathcal{X}^{\prime}(t)$ for $t\in\lbrack{\tau_{k},\tau_{k+1})}$ is
calculated through
\begin{equation}
\mathcal{X}^{\prime}(t)=\mathcal{X}^{\prime}(\tau_{k}^{+})+\int_{\tau_{k}}%
^{t}\frac{d}{dt}\mathcal{X}^{\prime}(t)dt \label{Xprime_t}%
\end{equation}
Table \ref{eventlist} contains all possible \emph{endogenous} event types for
our hybrid system. To these, we add \emph{exogenous} events $\kappa_{i}$,
$i=1,...,M$, to allow for possible discontinuities (jumps) in the random
processes $\{\sigma_{i}(t)\}$ which affect the sign of $\sigma_{i}(t)-\mu
_{ij}P_{ij}(t)$ in (\ref{Xdot}). We will use the notation $e(\tau_{k})$ to denote the
event type occurring at $t=\tau_{k}$ with $e(\tau_{k})\in E$, the event set
consisting of all endogenous and exogenous events. Finally, we make the
following assumption which is needed in guaranteeing the unbiasedness of the
IPA gradient estimates: $({\bf A6})$ Two events occur at the same time w.p. $0$ unless one is directly caused by the other.
\subsubsection{Objective Function Gradient}
The sample function gradient $\nabla\mathcal{L}(\Theta,T)$ needed in
(\ref{SAalgo}) is obtained from (\ref{ParamOptim}) assuming a total of $K$
events over $[0~T]$ with $\tau_{\!_{K+1}}=T$ and $\tau_{0}=0$:
\begin{equation}
\resizebox{1.0 \columnwidth}{!}{$ \begin{split} &\nabla \mathcal{L}(\Theta,T; {\bf X}(\Theta; 0)))=\frac{1}{T}\nabla\Big[\int_0^T\Big(\alpha \mathcal{L}_1(\Theta,t)-(1-\alpha)\mathcal{L}_2(\Theta,t)+\mathcal{L}_3(\Theta,t)\Big)dt\Big] \\ &\quad+\nabla \mathcal{L}_f(\Theta,T)\\ &=\frac{1}{T}\nabla\Big[\sum\limits_{k=0}^K\int_{\tau_k}^{\tau_{k+1}}\Big(\alpha \mathcal{L}_1(\Theta,t)-(1-\alpha)\mathcal{L}_2(\Theta,t)+\mathcal{L}_3(\Theta,t)\Big)dt\Big] \\&\quad+\nabla\mathcal{L}_f(\Theta,T)\\ &=\frac{1}{T}\Big[\sum\limits_{k=0}^K\Big(\alpha\Big(\int_{\tau_k}^{\tau_{k+1}}\nabla \mathcal{L}_1(\Theta,t)dt+\mathcal{L}_1(\Theta,\tau_{k+1})\tau^{\prime}_{k+1}-\mathcal{L}_1(\Theta,\tau_{k})\tau^{\prime}_{k}\Big)\\ &\quad-(1-\alpha)\Big(\int_{\tau_k}^{\tau_{k+1}}\nabla \mathcal{L}_2(\Theta,t)dt+\mathcal{L}_2(\Theta,\tau_{k+1})\tau^{\prime}_{k+1}-\mathcal{L}_2(\Theta,\tau_{k})\tau^{\prime}_{k}\Big)\\ &\quad+\Big(\int_{\tau_k}^{\tau_{k+1}}\nabla \mathcal{L}_3(\Theta,t)dt+\mathcal{L}_3(\Theta,\tau_{k+1})\tau^{\prime}_{k+1}-\mathcal{L}_3(\Theta,\tau_{k})\tau^{\prime}_{k}\Big)\Big]+\nabla \mathcal{L}_f(\Theta,T)\\ &=\frac{1}{T}\Big[\sum\limits_{k=0}^K\int_{\tau_k}^{\tau_{k+1}}\Big(\alpha\nabla \mathcal{L}_1(\Theta,t)dt-(1-\alpha)\nabla \mathcal{L}_2(\Theta,t)dt+\nabla \mathcal{L}_3(\Theta,t)dt\Big)\Big]\\&\quad+\nabla \mathcal{L}_f(\Theta,T) \end{split}$}
\label{gradJ}%
\end{equation}

The last step follows from the continuity of the state variables which causes
adjacent limit terms in the sum to cancel out. Therefore, $\nabla
\mathcal{L}(\Theta,T)$ does not have any direct dependence on any $\tau
_{k}^{\prime}$; this dependence is indirect through the state derivatives
involved in the four individual gradient terms. Referring to (\ref{J1}), the
first term involves $\nabla\mathcal{L}_{1}(\Theta,t)$ which is as a sum of
$X_{i}^{\prime}(t)$ derivatives. Similarly, $\nabla\mathcal{L}_{2}%
(\Theta,t)$ is a sum of $Y_{i}^{\prime}(t)$ derivatives and $\nabla
\mathcal{L}_{f}(\Theta,T)$ requires only $Z_{ij}^{\prime}(T)$. The third
term, $\nabla\mathcal{L}_{3}(\Theta,t)$, requires derivatives of $I_{j}(t)$
in (\ref{Ij}) which depend on the derivatives of the max function in
\eqref{Dplus} and the agent state derivatives $s_{j}^{^{\prime}}(t)$ with
respect to $\Theta$. Possible discontinuities in these derivatives occur when any
of the last four events in Table \ref{eventlist} takes place.

In summary, the evaluation of \eqref{gradJ} requires the state derivatives
$X_{i}^{\prime}(t)$, $Z_{ij}^{\prime}(t)$, $Y_{i}^{\prime}(t)$, and
$s_{j}^{^{\prime}}(t)$. The latter are easily obtained for any specific choice
of $f$ and $g$ in (\ref{param_traj}) and are shown in Appendix \ref{AppEllipse}. The former
require a rather laborious use of (\ref{tauprime})-(\ref{Xprime}) which,
however, reduces to a simple set of state derivative dynamics as shown next.

\emph{Proposition 1}: After an event occurrence at $t=\tau_{k}$, the state
derivatives $X_{i}^{\prime}(\tau_{k}^{+})$, $Y_{i}^{\prime}(\tau_{k}^{+})$,
$Z_{ij}^{\prime}(\tau_{k}^{+})$,  with respect to the controllable parameter
$\Theta$ satisfy the following:%
\[
X_{i}^{\prime}(\tau_{k}^{+})=\left\{
\begin{array}
[c]{ll}%
0 & \text{if }e(\tau_{k})=\xi_{i}^{0}\\
X_{i}^{\prime}(\tau_{k}^{-})-\mu_{il}(t)P_{il}(\tau_{k}){\tau_{k}^{^{\prime}}}
& \text{if }e(\tau_{k})=\delta_{ij}^{+}\\
X_{i}^{\prime}(\tau_{k}^{-}) & \text{otherwise}%
\end{array}
\right.
\]
where $l\neq j$ with $P_{il}(\tau_{k})>0$ if such $l$ exists and ${\tau
_{k}^{^{\prime}}=}\frac{\partial D_{ij}(s_{j})}{\partial s_{j}}\frac{\partial
s_{j}}{\partial\Theta}\left(  \frac{\partial D_{ij}(s_{j})}{\partial s_{j}%
}\dot{s}_{j}(\tau_{k})\right)  ^{-1}$.
\begin{align*}
Y_{i}^{\prime}(\tau_{k}^{+}) &  =\left\{
\begin{array}
[c]{ll}%
Y_{i}^{\prime}(\tau_{k}^{-})+Z_{ij}^{\prime}(\tau_{k}^{-}) & \text{if }%
e(\tau_{k})=\zeta_{ij}^{0}\\
Y_{i}^{\prime}(\tau_{k}^{-}) & \text{otherwise}%
\end{array}
\right.  \\
Z_{ij}^{\prime}(\tau_{k}^{+}) &  =\left\{
\begin{array}
[c]{ll}%
0 & \text{if }e(\tau_{k})=\zeta_{ij}^{0}\\
Z_{ij}^{\prime}(\tau_{k}^{-})+X_{i}^{\prime}(\tau_{k}^{-}) & \text{if }%
e(\tau_{k})=\xi_{i}^{0}\\
Z_{ij}^{\prime}(\tau_{k}^{-}) & \text{otherwise}%
\end{array}
\right.
\end{align*}
where $e(\tau_{k})=\xi_{i}^{0}$ occurs when  $j$ is connected to target $i$.

\emph{Proof}: See \eqref{Xprime0}, \eqref{xiplus}, \eqref{Xplusjump}, \eqref{Yprime}, \eqref{yrplus}, \eqref{zijplus}, \eqref{Zprime0}, \eqref{Zprime01} in Appendix \ref{IPAapp}.

This result shows that only three of the events in $E$ can actually cause
discontinuous changes to the state derivatives. Further, note that
$X_{i}^{\prime}(t)$ is reset to zero after a $\xi_{i}^{0}$ event. Moreover,
when such an event occurs, note that $Z_{ij}^{\prime}(t)$ is coupled to
$X_{i}^{\prime}(t)$. Similarly for $Z^{\prime}_{ij}(t)$ and $Y^{\prime}_{i}(t)$ when event
$\zeta_{ij}^{0}$ occurs, showing that perturbations in $\Theta$ can only
propagate to an adjacent queue when that queue is emptied.

\emph{Proposition 2}: The state derivatives $X_{i}^{\prime}(\tau_{k+1}^{-})$,
$Y_{i}^{\prime}(\tau_{k+1}^{-})$ with respect to the controllable parameter
$\Theta$ satisfy the following after an event occurrence at $t=\tau_{k}$:%

\begin{align*}
X_{i}^{\prime}(\tau_{k+1}^{-}) &  =\left\{
\begin{array}
[c]{ll}%
0 & \text{if }e(\tau_{k})=\xi_{i}^{0}\\
X_{i}^{\prime}(\tau_{k}^{+})-\int_{\tau_{k}}^{\tau_{k+1}}\mu_{ij}%
P_{ij}^{\prime}(u)du & \text{otherwise}%
\end{array}
\right.  \\
Y_{i}^{\prime}(\tau_{k+1}^{-}) &  =Y_{i}^{\prime}(\tau_{k}^{+})+\int_{\tau
_{k}}^{\tau_{k+1}}\beta_{i}^{\prime}(u)du
\end{align*}
where $j$ is such that $P_{ij}(t)>0$, $t\in\lbrack{\tau_{k},\tau_{k+1})}$.

\emph{Proof}: See \eqref{Xprimet1}, \eqref{dxdt} and \eqref{dydt} in Appendix
\ref{IPAapp}.

\emph{Proposition 3}: The state derivatives $Z_{ij}^{\prime}(\tau_{k+1}^{+})$
with respect to the controllable parameter $\Theta$ satisfy the following
after an event occurrence at $t=\tau_{k}$:\\
\emph{i}- If $j$ is connected to target $i$,%
\[
Z_{ij}^{\prime}(\tau_{k+1}^{-})=\left\{
\begin{array}
[c]{ll}%
Z_{ij}^{\prime}(\tau_{k}^{+}) \qquad\qquad\text{if }e(\tau_{k})=\xi_{i}^{0},\text{
}\zeta_{ij}^{0}\text{ or }\delta_{ij}^{+}\\
Z_{ij}^{\prime}(\tau_{k}^{+})+\int_{\tau_{k}}^{\tau_{k+1}}\mu_{ij}%
P_{ij}^{\prime}(u)du  \quad\text{otherwise}%
\end{array}
\right.
\]
\emph{ii}- If $j$ is connected to $B$ with $Z_{ij}(\tau_{k})>0$,
\[
Z_{ij}^{\prime}(\tau_{k+1}^{-})=Z_{ij}^{\prime}(\tau_{k}^{+})-\int_{\tau_{k}%
}^{\tau_{k+1}}\beta_{ij}P_{Bj}^{\prime}(u)du
\]
\emph{iii}- Otherwise, $Z_{ij}^{\prime}(\tau_{k+1}^{-})=Z_{ij}^{\prime}(\tau_{k}^{+})$.

\emph{Proof}: See \eqref{dzdt}, \eqref{dzdt1}, \eqref{Zprimet0} and \eqref{ZprimetB} in Appendix
\ref{IPAapp}.

\begin{corollary}\label{corollary1} The state derivatives $X_{i}^{\prime}(t)$, $Z_{ij}^{\prime
}(t)$, $Y_{i}^{\prime}(t)$ with respect to the controllable parameter
$\Theta$ are independent of the random data arrival processes $\{\sigma
_{i}(t)\}$, $i=1,\ldots,M$.
\end{corollary}
\emph{Proof}: Follows directly from the three Propositions.

There are a few important consequences of these results. First, as the
Corollary asserts, one can apply IPA regardless of the characteristics of the
random processes $\{\sigma_{i}(t)\}$. This robustness property does not
mean that these processes do not affect the values of the $X_{i}^{\prime}(t)$,
$Z_{ij}^{\prime}(t)$, $Y_{i}^{\prime}(t)$; this happens through the values of
the event times ${\tau_{k}}$, $k=1,2,\ldots$, which are observable and enter
the computation of these derivatives as seen above. Second, the IPA estimation
process is event-driven: $X_{i}^{\prime}(\tau_{k}^{+})$, $Y_{i}^{\prime}%
(\tau_{k}^{+})$, $Z_{ij}^{\prime}(\tau_{k}^{+})$ are evaluated at event times
and then used as initial conditions for the evaluations of $X_{i}^{\prime}%
(\tau_{k+1}^{-})$, $Y_{i}^{\prime}(\tau_{k+1}^{-})$, $Z_{ij}^{\prime}%
(\tau_{k+1}^{-})$ along with the integrals appearing in Propositions 2,3 which
can also be evaluated at $t=\tau_{k+1}$. Consequently, this approach is
scalable in the number of events in the system as the number of agents and
targets increases. Third, despite the elaborate derivations in the Appendix,
the actual implementation reflected by the three Propositions is simple.
Finally, returning to (\ref{gradJ}), note that the integrals involving
$\nabla\mathcal{L}_{1}(\Theta,t)$, $\nabla\mathcal{L}_{2}(\Theta,t)$ are
directly obtained from $X_{i}^{\prime}(t)$, $Y_{i}^{\prime}(t)$, the integral
involving $\nabla\mathcal{L}_{3}(\Theta,t)$ is obtained from straightforward
differentiation of (\ref{Ij}), and the final term is obtained from
$Z_{ij}^{\prime}(T)$.

\subsubsection{Objective Function Optimization}
This is carried out using (\ref{SAalgo}) with an appropriate step size sequence.
\subsection{Elliptical Trajectories}
Elliptical trajectories are described by their center coordinates, minor and
major axes and orientation. Agent $j$'s position $s_{j}(t)=[s_{j}^{x}%
(t),s_{j}^{y}(t)]$ follows the general parametric equation of the ellipse:
\begin{equation}%
\begin{array}
[c]{ll}%
s_{j}^{x}(t)= & A_{j}+a_{j}\cos\rho_{j}(t)\cos\phi_{j}-b_{j}\sin\rho
_{j}(t)\sin\phi_{j}\\
s_{j}^{y}(t)= & B_{j}+a_{j}\cos\rho_{j}(t)\sin\phi_{j}+b_{j}\sin\rho
_{j}(t)\cos\phi_{j}%
\end{array}
\end{equation}
Here, $\Theta_{j}=[A_{j},B_{j},a_{j},b_{j},\phi_{j}]$ where $A_{j},B_{j}$
are the coordinates of the center, $a_{j}$ and $b_{j}$ are the major and minor
axis respectively while $\phi_{j}\in\lbrack0,\pi)$ is the ellipse orientation
which is defined as the angle between the $x$ axis and the major axis of the
ellipse. The time dependent parameter $\rho_{j}(t)$ is the eccentric anomaly
of the ellipse. Since the agent is moving with constant speed of 1 on this
trajectory from \eqref{OptimU}, we have $\dot{s}_{j}^{x}(t)^{2}+\dot{s}_{j}%
^{y}(t)^{2}=1$ which gives
\begin{equation}
\resizebox{0.99\columnwidth}{!}{$ \dot\rho_j(t)=\left[ \begin{array}{l l}&\Big(a\sin\rho_j(t)\cos\phi_j+b_j\cos\rho_j(t)\sin\phi_j\Big)^2\\ &\quad+\Big(a\sin\rho_j(t)\sin\phi_j-b_j\cos\rho_j(t)\cos\phi_j\Big)^2 \end{array}\right]^{-\frac{1}{2}}$}
\end{equation}
In the data harvesting problem, trajectories that do not pass through the base
are inadmissible since there is no delivery of data. Therefore, we add a
constraint to force the ellipse to pass through $w_{\!_{B}}=[w_{\!_{B}}%
^{x},w_{\!_{B}}^{y}]$ where:
\begin{equation}%
\begin{split}
w_{\!_{B}}^{x}= &  A_{j}+a_{j}\cos\rho_{j}(t)\cos\phi_{j}-b_{j}\sin\rho
_{j}(t)\sin\phi_{j}\\
w_{\!_{B}}^{y}= &  B_{j}+a_{j}\cos\rho_{j}(t)\sin\phi_{j}+b_{j}\sin\rho
_{j}(t)\cos\phi_{j}%
\end{split}
\end{equation}
Using the fact that $\sin^{2}\rho(t)+\cos^{2}\rho(t)=1$ we define a quadratic
constraint term added to $J(\Theta,T;\mathbf{X}(\Theta,0))$ with a
sufficiently large multiplier. This can ensure the optimal path passes
through the base location. We define $\mathcal{C}_{j}(\Theta_{j})$ which
appears in \eqref{ParamOptim_e}:
\begin{equation}
\mathcal{C}_{j}(\Theta_{j})=\big(1-f_{j}^{1}\cos^{2}\phi_{j}-f_{j}^{2}%
\sin^{2}\phi_{j}-f_{j}^{3}\sin2\phi_{j}\big)^{2}%
\end{equation}
where $f_{j}^{1}=\big(\frac{w_{\!_{B}}^{x}-A_{j}}{a_{j}}\big)^{2}%
+\big(\frac{w_{\!_{B}}^{y}-B_{j}}{b_{j}}\big)^{2}$, $f_{j}^{2}=\big(\frac
{w_{\!_{B}}^{x}-A_{j}}{b_{j}}\big)^{2}+\big(\frac{w_{\!_{B}}^{y}-B_{j}}{a_{j}%
}\big)^{2}$, $f_{j}^{3}=\frac{(b_{j}^{2}-a_{j}^{2})(w_{\!_{B}}^{x}%
-A_{j})(w_{\!_{B}}^{y}-B_{j})}{a_{j}^{2}b_{j}^{2}}$.

Multiple visits to the base may be needed during the mission time $[0,T]$. We
can capture this by allowing an agent trajectory to consist of a sequence of
admissible ellipses. For each agent, we define $\mathcal{E}_{j}$ as the number
of ellipses in its trajectory. The parameter vector $\Theta_{j}^{\kappa}$
with $\kappa=1,\dots,\mathcal{E}_{j}$, defines the $\kappa^{th}$ ellipse in agent
$j$'s trajectory and $\mathcal{T}_{j}^{\kappa}$ is the time that agent $j$
completes ellipse $\kappa$. Therefore, the location of each agent is described
through $\kappa$ during $[\mathcal{T}_{j}^{\kappa-1},\mathcal{T}_{j}^{\kappa
}]$ where $\mathcal{T}_{j}^{0}=0$. Since we cannot optimize over all possible
$\mathcal{E}_{j}$ for all agents, an iterative process needs to be performed
in order to find the optimal number of segments in each agent's trajectory. At
each step, we fix $\mathcal{E}_{j}$ and find the optimal trajectory with that
many segments. The process is stopped once the optimal trajectory with
$\mathcal{E}_{j}$ segments is no better than the optimal one with
$\mathcal{E}_{j}-1$ segments (obviously, this is not a globally optimal
solution). We can now formulate the parametric optimization problem
$\mathbf{P2_{e}}$ where $\Theta_{j}=[\Theta_{j}^{1},\dots,\Theta
_{j}^{\mathcal{E}_{j}}]$ and $\Theta=[\Theta_{1},\dots,\Theta_{N}]$:
\begin{equation}
\resizebox{0.99 \columnwidth}{!}{$ \begin{split} \min\limits_{\Theta\in F_{\Theta}} J_e=&\frac{1}{T}\int_0^T\Big[\alpha J_1(\Theta,t)-(1-\alpha)J_2(\Theta,t)+J_3(\Theta,t)\Big]dt\\&+M_C\sum\limits_{j=1}^N\mathcal{C}_j(\Theta_j)+J_f(\Theta,T) \end{split}$}\label{ParamOptim_e}%
\end{equation}

where $M_{C}$ is a large multiplier. The evaluation of $\nabla\mathcal{C}_{j}$ is straightforward and does not depend on any event.
(Details are shown in Appendix \ref{AppEllipse}).
\subsection{Fourier Series Trajectories}
The elliptical trajectories are limited in shape and may not be able to cover
many targets in a mission space. Thus, we next parameterize the trajectories
using a Fourier series representation of closed curves \cite{Zahn1972}. Using
a Fourier series function for $f$ and $g$ in \eqref{param_traj}, agent $j$'s
trajectory can be described as follows with base frequencies $f_j^{x}$ and
$f_j^{y}$:
\begin{equation}%
\begin{array}
[c]{ll}%
s_{j}^{x}(t)= & a_{0,j}+\displaystyle\sum_{n=1}^{\Gamma_{j}^{x}}a_{n,j}%
\sin(2\pi nf_{j}^{x}\rho_{j}(t)+\phi_{n,j}^{x})\\
s_{j}^{y}(t)= & b_{0,j}+\displaystyle\sum_{n=1}^{\Gamma_{j}^{y}}b_{n,j}%
\sin(2\pi nf_{j}^{y}\rho_{j}(t)+\phi_{n,j}^{y})
\end{array}
\label{fouriertraj}
\end{equation}
The parameter $\rho(t)\in\lbrack0,2\pi]$, similar to elliptical trajectories,
represents the position of the agent along the trajectory. In this case,
forcing a Fourier series curve to pass through the base is easier. For
simplicity, we assume a trajectory to start at the base and set $s_{j}%
^{x}(0)=w_{\!_{B}}^{x}$, $s_{j}^{y}(0)=w_{\!_{B}}^{y}$. Assuming $\rho(0)=0$,  with no loss of generality, we can calculate the zero frequency terms by means
of the remaining parameters:
\begin{equation}
\resizebox{0.98 \columnwidth}{!}{$ a_{0,j}=w^x_{\!_B}-\displaystyle\sum_{n=1}^{\Gamma_j^x}a_{n,j}\sin(\phi^x_{n,j}), b_{0,j}=w^y_{\!_B}-\displaystyle\sum_{n=1}^{\Gamma_j^y}b_{n,j}\sin(\phi^y_{n,j})$}\label{fourier_baseconst}%
\end{equation}

The parameter vector for agent $j$ is $\Theta_{j}=[f_{j}^{x},a_{0,j}%
,\ldots,a_{\Gamma_{j}^{x}},b_{0,j},\ldots,b_{\Gamma_{j}^{y}},\phi_{1,j}%
,\ldots,\phi_{\Gamma_{j}^{x}},\xi_{1,j},\ldots,\xi_{\Gamma_{j}^{y}}]$ and
$\Theta=[\Theta_{1},\ldots,\Theta_{N}]$. Note that the shape of the
curve is fully represented by the ratio $f_{j}^{x}/f_{j}^{y}$ so one of these
can be kept constant. For the Fourier trajectories, the fact that $\mathbf{u}_{j}^{\ast}=1$
allows us to calculate $\dot{\rho}_{j}(t)$ as follows:
\begin{equation}
\resizebox{0.98 \columnwidth}{!}{$\dot \rho_j(t)=\frac{1}{2\pi}\left[\begin{array}{l l}&\Bigg(f_j^x\displaystyle\sum_{n=1}^{\Gamma_j^x}a_{n,j}n\sin(2\pi f_j^x\rho_j(t)+\phi^x_{n,j})\Bigg)^2\\ &\quad+\Bigg(f_j^y\displaystyle\sum_{n=1}^{\Gamma_j^x}b_{n,j}n\sin(2\pi f_j^y\rho_j(t)+\phi^y_{n,j})\Bigg)^2\end{array}\right]^{-1/2}$}
\end{equation}

Problem $\mathbf{P2_{f}}$ is the same as $\mathbf{P2}$
but there are no additional constraints in this case:
\begin{equation}
\resizebox{1.0 \columnwidth}{!}{$ \min\limits_{\Theta\in F_{\Theta}} J_f=\frac{1}{T}\int_0^T\Big(\alpha J_1(t)-(1-\alpha)J_2(t)+J_3(t)\Big)+J_f(T)$}\label{ParamOptim_f}%
\end{equation}
\section{Numerical Results}
\label{Numerical}
\begin{figure}
  \centering
  \includegraphics[scale=0.3]{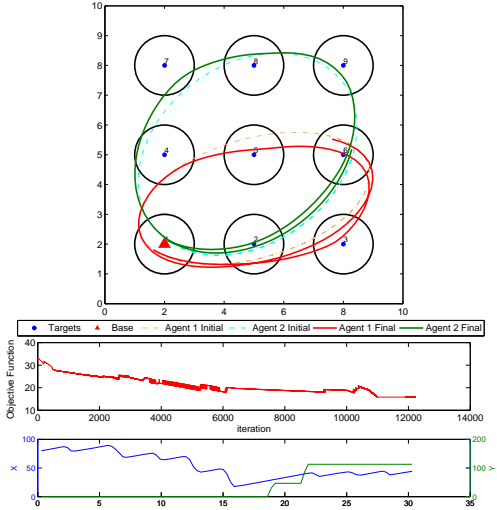}\\
  \caption{8-targets, 2-agents, TPBVP trajectories (T=30) $J^*=15.82$}\label{TPBVP9}
\end{figure}

\begin{figure}
  \centering
  \includegraphics[scale=0.3]{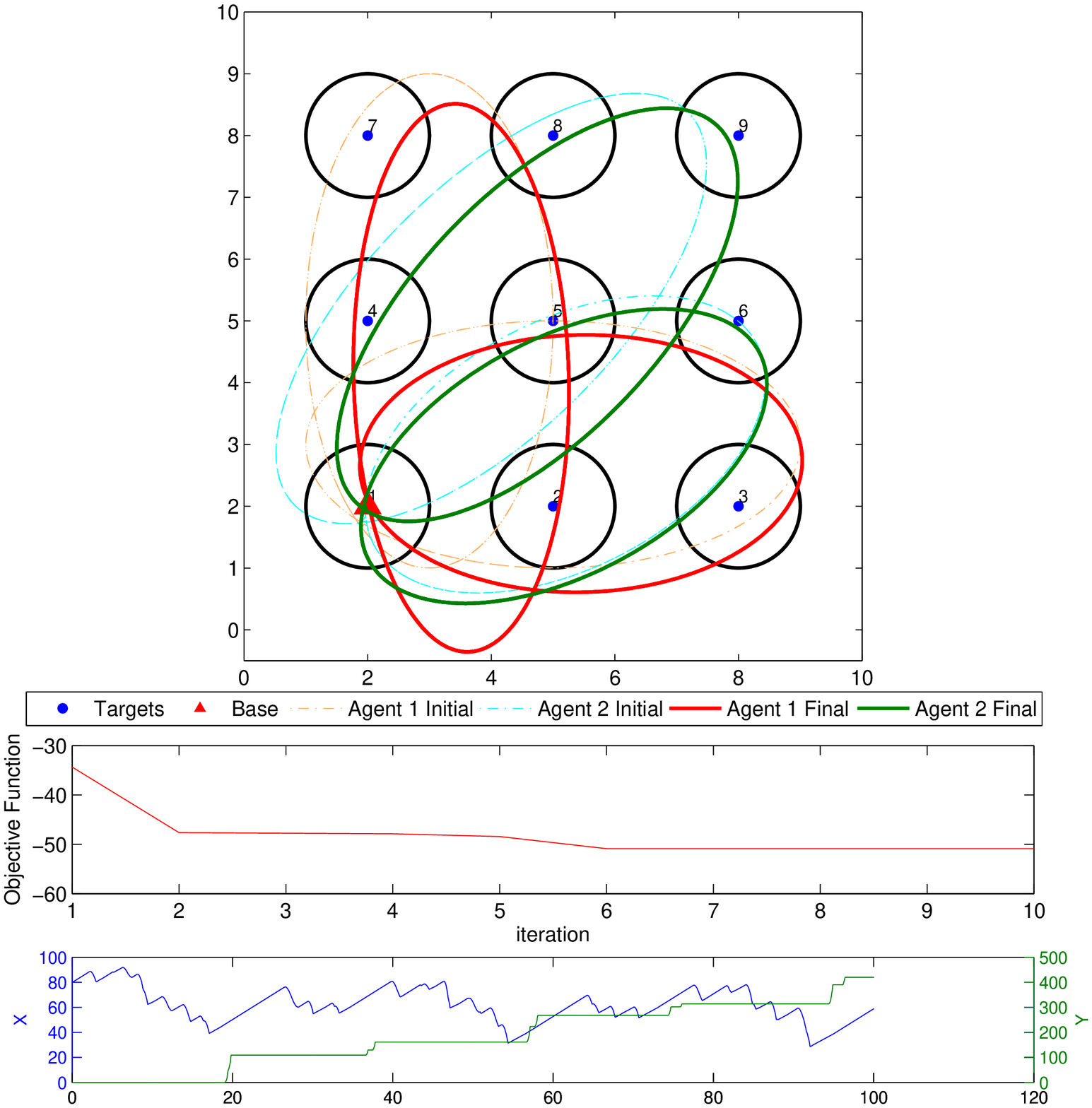}\\
  \caption{8-targets, 2-agents, Elliptical trajectories (T=100) $J^*=-50.9$}\label{Ellipse9}
\end{figure}

\begin{figure}
  \centering
  \includegraphics[scale=0.3]{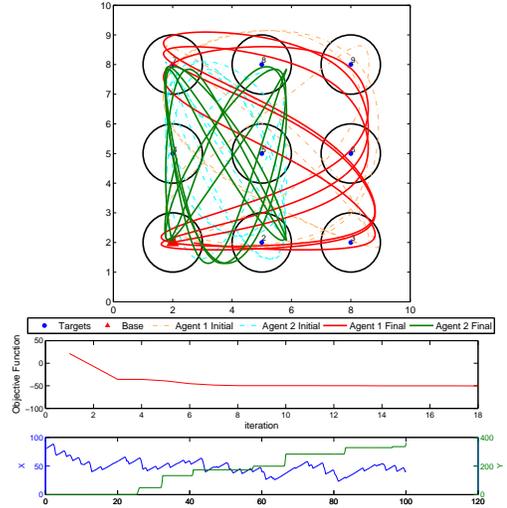}\\
  \caption{8-targets, 2-agents, Fourier series trajectories (T=100) $J^*=-50.18$}\label{Fourier9}
\end{figure}

In this section numerical results are presented to illustrate our approach. We consider 8 targets, 2 agents and a base as shown in Fig. \ref{TPBVP9}. First, we assume deterministic arrival process with $\sigma_i=0.5$ for all $i$. For \eqref{Pij} and \eqref{PB} we have used $p(w,v) = \max(0, 1-\frac{D(w,v)}{r})$ where $r$ is the corresponding value of $r_{ij}$ or $r_{\!_{Bj}}$. We have $\mu_{ij}=50$ and $\beta_{ij}=500$ for all $i$ and $j$. Other parameters used are $\alpha=0.5$, $r_{ij}=r_{\!_{Bj}}=1$, $M_I=1$ and $T=100$ except for the TPBVP case where $T=30$. In Fig. \ref{TPBVP9} results of the TPBVP are shown which depend heavily on the initial trajectory and this is the best result among several initializations. These results are after 10,000 iterations of the TPBVP solver. In Fig. \ref{Ellipse9} the results are shown for the (locally) optimal trajectory with two ellipses in each agent's trajectory ($\mathcal{E}_{j}=2$) and in Fig. \ref{Fourier9} for a Fourier series representation with 5 terms in \eqref{fouriertraj}. Both methods converge in few iterations with each iteration taking less than a few seconds. We use the Armijo rule to update the step-size in each iteration. The average queue length at targets for TPBVP, Ellipse with $\mathcal{E}_j=2$ and Fourier series are 52.13, 49.23 and 62.03 respectively. Whereas The average throughput for the three trajectories is 3.76, 4.2, 3.56 respectively. Although the example is a very symmetric configuration, the benefit of the Fourier series trajectories shows when the targets are randomly positioned. Then, initializing the TPBVP becomes a very hard task and ellipses cannot fit all targets.

\begin{figure}
  \centering
  \includegraphics[scale=0.3]{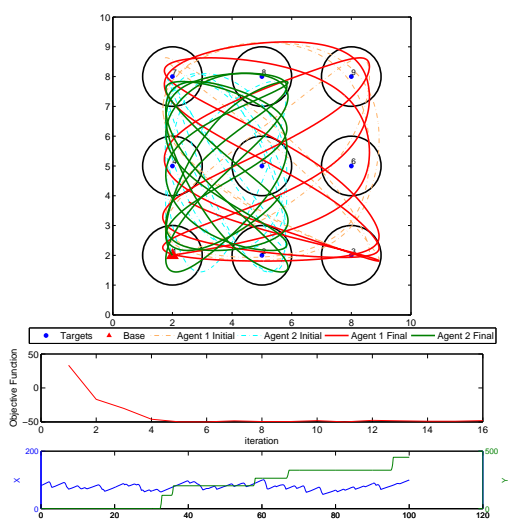}\\
  \caption{8-targets, 2-agents, Random data processes - Fourier series trajectories $J^*=-48.05$}\label{Fourier9stoch}
\end{figure}

Based on Corollary \ref{corollary1} our results are independent of the underlying random processes $\{\sigma_i(t)\}$. To verify this property, we model the exact same problem with a uniform distribution for $\sigma_i(t)$ as $U[0.1,0.9]$. Note that we keep $E[\sigma_i(t)]=0.5$, the same rate as in the deterministic setting. At each iteration we generate a random sample path using the random process with $\sigma_i(t)\sim U[0.1,0.9]$. The Fourier series trajectories for this stochastic optimization problem are shown in Fig. \ref{Fourier9stoch} with $J^*=-48.05$ compared to $J^*=-50.18$. The objective function converges almost as quickly but with some oscillations as expected.
\section{Conclusions}
\label{Conclusions}
We have developed a new method for trajectory planning in the data harvesting problem. An optimal control formulation provides initial insights for the solution, but it is computationally intractable, especially in the case where the data generating processes are stochastic. We propose an agent trajectory parameterization in terms of general function families which are optimized on line through the use of IPA. Explicit results are provided for the case of elliptical and Fourier series trajectories. We have shown robustness of the solution with respect to stochastic data generation processes by considering stochastic data arrivals at targets. Natural next steps include constraining trajectories to urban setting obstacles in the mission space.

\begin{appendices}
\section{Elliptical Trajectories}
\label{AppEllipse}
In order to calculate the IPA derivatives we need to have the derivative of state variable with respect to all the parameter vector $\Theta_j=[A_j,B_j,a_j,b_j,\phi_j]$ for all agents $j$. These derivatives do not depend on the events happening in the system since the trajectories of agents are fixed at each iteration. For now we assume $\mathcal{E}_j=1$ for all $j=1,\dots,N$ hence, we drop the superscript. We have:
\begin{equation}
\frac{\partial s^x_j}{\partial A_j}=1,\qquad \frac{\partial s^x_j}{\partial B_j}=0
\label{dsxdAB}
\end{equation}
\begin{equation}
\frac{\partial s^x_j}{\partial a_j}=\cos\rho_j(t)\cos\phi_j,\qquad\frac{\partial s^x_j}{\partial b_j}=-\sin\rho_j(t)\sin\phi_j
\label{dsxdab}
\end{equation}
\begin{equation}
\frac{\partial s^x_j}{\partial \phi_j}=-a_j\cos\rho_j(t)\sin\phi_j-b_j\sin\rho_j(t)\cos\phi_j
\label{dsxdphi}
\end{equation}
\begin{equation}
\frac{\partial s^y_j}{\partial A_j}=0,\qquad \frac{\partial s^y_j}{\partial B_j}=1
\label{dsydAB}
\end{equation}
\begin{equation}
\frac{\partial s^y_j}{\partial a_j}=\cos\rho_j(t)\sin\phi_j,\qquad\frac{\partial s^y_j}{\partial b_j}=\sin\rho_j(t)\cos\phi_j
\label{dsydab}
\end{equation}
\begin{equation}
\frac{\partial s^y_j}{\partial \phi_j}=a_j\cos\rho_j(t)\cos\phi_j-b_j\sin\rho_j(t)\sin\phi_j
\label{dsydphi}
\end{equation}
Also the time derivative of the position state variables are calculated as below:
\begin{equation}
  \dot s_j^x(t)=-a_j\dot\rho_j(t)\sin\rho_j(t)\cos\phi_j+b_j\dot\rho_j(t)\cos\rho_j(t)\sin\phi_j
  \label{dsxdt}
\end{equation}
\begin{equation}
  \dot s_j^y(t)=-a_j\dot\rho_j(t)\sin\rho_j(t)\sin\phi_j+b_j\dot\rho_j(t)\cos\rho_j(t)\cos\phi_j
  \label{dsydt}
\end{equation}
The gradient of the last term in the $J_e$ in \eqref{ParamOptim_e} needs to be calculated separately. We have for $j \ne l$, $\frac{\partial \mathcal{C}_j}{\partial \Theta_l}=0$ and for $j=l$:
\begin{equation*}
\resizebox{1.0 \columnwidth}{!}{$
  \frac{\partial \mathcal{C}_j}{\partial A_j} =2\mathcal{C}_j\big(-\cos^2\phi_j\frac{\partial f^1_j}{\partial A_j}-\sin^2\phi_j\frac{\partial f^2_j}{\partial A_j}-\sin2\phi_j\frac{\partial f^3_j}{\partial A_j}\big)$}
\end{equation*}
\begin{equation*}
\resizebox{1.0 \columnwidth}{!}{$
  \frac{\partial \mathcal{C}_j}{\partial B_j} =2\mathcal{C}_j\big(-\cos^2\phi_j\frac{\partial f^1_j}{\partial B_j}-\sin^2\phi_j\frac{\partial f^2_j}{\partial B_j}-\sin2\phi_j\frac{\partial f^3_j}{\partial B_j}\big)$}
\end{equation*}
\begin{equation*}
\resizebox{1.0 \columnwidth}{!}{$
  \frac{\partial \mathcal{C}_j}{\partial a_j} =2\mathcal{C}_j\big(-\cos^2\phi_j\frac{\partial f^1_j}{\partial a_j}-\sin^2\phi_j\frac{\partial f^2_j}{\partial a_j}-\sin2\phi_j\frac{\partial f^3_j}{\partial a_j}\big)$}
\end{equation*}
\begin{equation*}
\resizebox{1.0 \columnwidth}{!}{$
  \frac{\partial \mathcal{C}_j}{\partial b_j} =2\mathcal{C}_j\big(-\cos^2\phi_j\frac{\partial f^1_j}{\partial b_j}-\sin^2\phi_j\frac{\partial f^2_j}{\partial b_j}-\sin2\phi_j\frac{\partial f^3_j}{\partial b_j}\big)$}
\end{equation*}
\begin{equation*}
\resizebox{0.78 \columnwidth}{!}{$
  \frac{\partial \mathcal{C}_j}{\partial \phi_j}
  =2\mathcal{C}_j\big((f_j^1-f_j^2)\sin2\phi_j-2f_j^3\cos2\phi_j\big)$}
\end{equation*}
where
\begin{equation*}\begin{split}
  \frac{\partial f^1_j}{\partial A_j}&=-2\Big(\frac{w_{\!_B}^x-A_j}{a_j^2}\Big),\quad
  \frac{\partial f^1_j}{\partial B_j}=-2\Big(\frac{w_{\!_B}^y-B_j}{b_j^2}\Big)\\
  \frac{\partial f^1_j}{\partial a_j}&=-2\Big(\frac{(w_{\!_B}^x-A_j)^2}{a_j^3}\Big),\quad
    \frac{\partial f^1_j}{\partial b_j}=-2\Big(\frac{(w_{\!_B}^y-B_j)^2}{b_j^3}\Big)\\
  \end{split}
\end{equation*}
\begin{equation*}\begin{split}
  \frac{\partial f^2_j}{\partial A_j}&=-2\Big(\frac{w_{\!_B}^x-A_j}{b_j^2}\Big),~
  \frac{\partial f^2_j}{\partial B_j}=-2\Big(\frac{w_{\!_B}^y-B_j}{a_j^2}\Big)\\
  \frac{\partial f^2_j}{\partial a_j}&=-2\Big(\frac{(w_{\!_B}^y-B_j)^2}{a_j^3}\Big),~
    \frac{\partial f^2_j}{\partial b_j}=-2\Big(\frac{(w_{\!_B}^x-A_j)^2}{b_j^3}\Big)\\
  \end{split}
\end{equation*}
  \begin{equation*}\begin{split}
  \frac{\partial f^3_j}{\partial A_j}&=-\Big(\frac{(b_j^2-a_j^2)(w_{\!_B}^y-B_j)}{a_j^2b_j^2}\Big)\\
  \frac{\partial f^3_j}{\partial B_j}&=-\Big(\frac{(b_j^2-a_j^2)(w_{\!_B}^x-A_j)}{a_j^2b_j^2}\Big)
    \end{split}
  \end{equation*}
    \begin{equation*}\begin{split}
  \frac{\partial f^3_j}{\partial a_j}&=-2\Big(\frac{(w_{\!_B}^x-A_j)(w_{\!_B}^y-B_j)}{a_j^3}\Big)\\
  \frac{\partial f^3_j}{\partial b_j}&=2\Big(\frac{(w_{\!_B}^x-A_j)(w_{\!_B}^y-B_j)}{b_j^3}\Big)
  \end{split}
  \end{equation*}
\section{Fourier Series Trajectories}
We calculate the position of agent $j$'s derivative with respect to all the Fourier parameters. The parameter vector is $\Theta_j=[f_j^x,a_{0,j},\ldots,a_{\Gamma_j^x},b_{0,j},\ldots,b_{\Gamma_j^y},\phi_{1,j},\ldots,\phi_{\Gamma_j^x},\xi_{1,j},\ldots,\xi_{\Gamma_j^y}]$. So we have:
\begin{equation}
\frac{\partial s^x_j}{\partial a_{0,j}}=1,\qquad \frac{\partial s^x_j}{\partial b_{0,j}}=0
\end{equation}
\begin{equation}
\frac{\partial s^x_j}{\partial a_{n,j}}=\sin(2\pi nf_j^x\rho_j(t)+\phi^x_{n,j}),\qquad\frac{\partial s^x_j}{\partial b_{n,j}}=0
\end{equation}
\begin{equation}
\frac{\partial s^x_j}{\partial \phi^x_{n,j}}=a_{n,j}\cos(2\pi nf_j^x\rho_j(t)+\phi^x_{n,j}) \qquad \frac{\partial s^x_j}{\partial \phi^y_{n,j}}=0
\end{equation}
\begin{equation}
\frac{\partial s^x_j}{\partial f_j^x}=2\pi\rho_j(t)\displaystyle\sum_{n=1}^{\Gamma_j^x}a_{n,j}n\cos(2\pi nf_j^x\rho_j(t)+\phi^x_{n,j}),
\end{equation}

\begin{equation}
\frac{\partial s^y_j}{\partial b_{0,j}}=1,\qquad \frac{\partial s^y_j}{\partial a_{0,j}}=0
\end{equation}
\begin{equation}
\frac{\partial s^y_j}{\partial b_{n,j}}=\sin(2\pi nf_j^y\rho_j(t)+\phi^y_{n,j}),\qquad\frac{\partial s^y_j}{\partial a_{n,j}}=0
\end{equation}
\begin{equation}
\frac{\partial s^y_j}{\partial \phi^y_{n,j}}=b_{n,j}\cos(2\pi nf_j^y\rho_j(t)+\phi^y_{n,j}) \qquad \frac{\partial s^y_j}{\partial \phi^x_{n,j}}=0
\end{equation}
\begin{equation}
\frac{\partial s^y_j}{\partial f_j^x}=0
\end{equation}
Also the time derivative of the position state variables are calculated as below:
\begin{equation}
  \dot s_j^x(t)=\dot\rho_j(t)\sum_{n=1}^{\Gamma_j^x}2\pi nf_j^xa_{n,j}\cos(2\pi nf_j^x\rho_j(t)+\phi^x_{n,j}),
\end{equation}
\begin{equation}
    \dot s_j^y(t)=\dot\rho_j(t)\sum_{n=1}^{\Gamma_j^y}2\pi nf_j^ya_{n,j}\cos(2\pi nf_j^y\rho_j(t)+\phi^x_{n,j}),
\end{equation}
\section{IPA events and derivatives}
\label{IPAapp}
In this section, we derive all event time derivatives and state
derivatives with respect to the controllable parameter $\Theta$ for each
event by applying the IPA equations.\\

1. {\bf Event $\xi_{i}^{0}$}: This event causes a transition from $X_{i}(t)>0$,
$t<\tau_{k}$ to $X_{i}(t)=0$, $t\geq\tau_{k}$. The switching function is
$g_{k}(\Theta,{\bf X})=X_{i}$ so $\frac{\partial g_{k}}{\partial X_{i}%
}=1$. From (\ref{tauprime}) and \eqref{Xdot}:
\begin{equation}%
\begin{split}
{\tau_{k}^{^{\prime}}}  &  =-\Big(\frac{\partial g_{k}}{\partial X_{i}}%
f_{k}(\tau_{k}^{-})\Big)^{-1}\Big(\frac{\partial g_{k}}{\partial\Theta
}+\frac{\partial g_{k}}{\partial X_{i}}{X_{i}^{\prime}(\tau_{k}^{-})}\Big)\\
&  =-\frac{X_{i}^{\prime}(\tau_{k}^{-})}{\sigma_{i}(\tau_{k})-\mu_{ij}%
P_{ij}(\tau_{k})}%
\end{split}
\label{tauprime1}%
\end{equation}
where agent $j$ is the one connected to $i$ at $t=\tau_{k}$ and we have used
the assumption that two events occur at the same time w.p. $0$, hence
$\sigma_{i}(\tau_{k}^{-})=\sigma_{i}(\tau_{k})$. From (\ref{dXdtprime}%
)-(\ref{Xprime}), since $\dot{X}_{i}(t)=0$, for $\tau_{k}\leq t<\tau_{k+1}$:
\begin{equation}
\frac{d}{dt}X_{i}^{\prime}(t)=\frac{\partial\dot{X}_{i}(t)}{\partial X_{i}%
(t)}X_{i}^{\prime}(t)+{\dot{X}_{i}^{\prime}(t)}=0 \label{Xprimet1}%
\end{equation}%
\begin{equation}%
\begin{split}
&  X_{i}^{\prime}(\tau_{k}^{+})=X_{i}^{\prime}(\tau_{k}^{-})+\Big[\Big(\sigma
_{i}(\tau_{k})-\mu_{ij}P_{ij}(\tau_{k})\Big)-0\Big]{\tau_{k}}^{\prime}\\
&  =X_{i}^{\prime}(\tau_{k}^{-})-\frac{X_{i}^{\prime}(\tau_{k}^{-}%
)\Big(\sigma_{i}(\tau_{k})-\mu_{ij}P_{ij}(\tau_{k})\Big)}{\sigma_{i}(\tau
_{k})-\mu_{ij}P_{ij}(\tau_{k})}=0
\end{split}
\label{Xprime0}%
\end{equation}
For $X_{r}(t)$, $r\neq i$, the dynamics of $X_{r}(t)$ in \eqref{Xdot} are
unaffected and we have:
\begin{equation}
X_{r}^{\prime}(\tau_{k}^{+})=X_{r}^{\prime}(\tau_{k}^{-}) \label{Xrprime0}%
\end{equation}
If $X_{r}(\tau_{k})>0$ and agent $l$ is connected to it, then%
\begin{equation}%
\begin{split}
&  \frac{d}{dt}X_{r}^{\prime}(t)=\frac{\partial\dot{X}_{r}(t)}{\partial
X_{r}(t)}X_{r}^{\prime}(t)+{\dot{X}_{r}^{\prime}(t)}\\
&  =\frac{\partial}{\partial\Theta}\Big(\sigma_{r}(t)-\mu_{rl}P_{rl}%
(\tau_{k})\Big)=-\mu_{rl}P_{rl}^{\prime}(t)
\end{split}
\label{dxdt}%
\end{equation}
and if ${X}_{r}(t)=0$ in $[\tau_k,\tau_{k+1}]$ or if no agents are connected to $i$, then and $\frac{d}{dt}X_{r}^{\prime}(t)=0$.\\
For $Y_{r}(t)$, $r=1,\dots,M$, the dynamics of $Y_{r}(t)$ in \eqref{Ydot} are not affected by the event $\xi_i^0$ at $\tau_k$, hence
\begin{equation}
Y_{r}^{\prime}(\tau_{k}^{+})=Y_{r}^{\prime}(\tau_{k}^{-}) \label{yrplus}%
\end{equation}
and since $\dot{Y}_{r}(t)=\beta_{r}(t)$, for $\tau_{k}\leq t<\tau_{k+1}$:
\begin{equation}%
\frac{d}{dt}Y_{r}^{\prime}(t)=\frac{\partial\dot{Y}_{r}(t)}{\partial Y_{r}%
(t)}Y_{r}^{\prime}(t)+{\dot{Y}_{r}^{\prime}(t)}=\beta_{r}^{\prime}(t)
\label{dydt}%
\end{equation}
For $Z_{ij}(t)$, we must have $Z_{ij}(\tau_{k})>0$ since $X_{i}(\tau_{k}%
^{-})>0$, hence $\tilde{\mu}_{ij}(\tau_{k}^{-})>0$ and from (\ref{Xprime}):
\begin{equation}%
\begin{split}
&  Z_{ij}^{\prime}(\tau_{k}^{+})=Z_{ij}^{\prime}(\tau_{k}^{-})+\Big[\dot
{Z}_{ij}(\tau_{k}^{-})-\dot{Z}_{ij}(\tau_{k}^{+})\Big]\tau_{k}^{\prime}\\
&  =Z_{ij}^{\prime}(\tau_{k}^{-})+\Big[\tilde{\mu}_{ij}(\tau_{k}^{-}%
)-\tilde{\mu}_{ij}(\tau_{k}^{+})\Big]P_{ij}(\tau_{k})\tau_{k}^{\prime}%
\end{split}
\label{Zprime}%
\end{equation}
Since $X_{i}(\tau_{k}^{-})>0$, from \eqref{mutilde} we have $\tilde{\mu}%
_{ij}(\tau_{k}^{-})=\mu_{ij}$. At $\tau_{k}^{+}$, $j$ remains connected to
target $i$ with $\tilde{\mu}_{ij}(\tau_{k}^{+})=\sigma_{i}(\tau_{k}%
^{+})/P_{ij}(\tau_{k})=\sigma_{i}(\tau_{k})/P_{ij}(\tau_{k})$ and we get%
\begin{equation}
\begin{split}
\label{Zprime01}
Z_{ij}^{\prime}(\tau_{k}^{+})  &  =Z_{ij}^{\prime}(\tau_{k}^{-})+\frac{-X_{i}^{\prime}(\tau
_{k}^{-})\Big[\mu
_{ij}P_{ij}(\tau_{k})-\sigma_{i}(\tau_{k})\Big]}{\sigma_{i}(\tau_{k})-\mu_{ij}P_{ij}(\tau_{k})}\\
&  =Z_{ij}^{\prime}(\tau_{k}^{-})+X_{i}^{\prime}(\tau_{k}^{-})
\end{split}
\end{equation}
From (\ref{dXdtprime}) for $\tau_{k}\leq t<\tau_{k+1}$:%
\begin{equation}%
\begin{split}
&  \frac{d}{dt}Z_{ij}^{\prime}(t)=\frac{\partial\dot{Z}_{ij}(t)}{\partial
Z_{ij}(t)}Z_{ij}^{\prime}(t)+\frac{\partial{\dot{Z}_{ij}(t)}}{\partial
\Theta}\\
&  =\frac{\partial{\dot{Z}_{ij}(t)}}{\partial\Theta}=\frac{\partial
}{\partial\Theta}\Big(\tilde{\mu}_{ij}(t)P_{ij}(t)-\beta_{ij}P_{\!_{Bj}%
}(t)\Big)
\end{split}
\label{dzdt}%
\end{equation}
Since $\tilde{\mu}_{ij}(t)=\sigma_{i}(t)/P_{ij}(t)$ for the agent which
remains connected to target $i$ after this event, it follows that
$\frac{\partial}{\partial\Theta}[\tilde{\mu}_{ij}(t)P_{ij}(t)]=0$. Moreover,
$P_{\!_{Bj}}(t)=0$ by our assumption that agents cannot be within range of the
base and targets at the same time and we get%
\begin{equation}
\frac{d}{dt}Z_{ij}^{\prime}(t)=0 \label{dzdt1}%
\end{equation}
Otherwise, for $r\neq j$, we have $\tilde{\mu}_{ir}(t)=0$ and we get:
\begin{equation}
\frac{d}{dt}Z_{ir}^{\prime}(t)=-\beta_{ir}P_{\!_{Br}}^{\prime}(t)
\end{equation}
Finally, for $Z_{rj}(t)$, $r\neq i$ we have $Z_{rj}^{\prime}(\tau_{k}%
^{+})=Z_{rj}^{\prime}(\tau_{k}^{-})$. If $Z_{rj}(t)=0$ in $[\tau_{k}%
,\tau_{k+1})$, then $\frac{d}{dt}Z_{rj}^{\prime}(t)=0$. Otherwise, we get
$\frac{d}{dt}Z_{rj}^{\prime}(t)$ from \eqref{dzdt} with $i$ replaced by $r$.\\

2. {\bf Event $\xi_{i}^{+}$}: This event causes a transition from $X_i(t)=0$, $t\le\tau_k$ to $X_i(t)>0$, $t > \tau_k$. Note that this transition can occur as an exogenous event when an empty queue $X_i(t)$ gets a new arrival in which case we simply have $\tau_k^{\prime}=0$ since the exogenous event is independent of the controllable parameters. In the endogenous case, however, we have the switching function $g_k(\Theta,{\bf X})=\sigma_i(t)- \mu_{ij}P_{ij}(t)$ in which agent $j$ is connected to target $i$ at $t=\tau_k$. Assuming $\frac{\partial s_j}{\partial \Theta}=\big[\frac{\partial s_j^x}{\partial \Theta}~\frac{\partial s_j^y}{\partial \Theta}\big]^{\top}$ and $\dot s_j=[\dot s_j^x~ \dot s_j^y]^{\top}$, from (\ref{tauprime}):
\begin{equation}
\begin{split}
  {\tau_k}^{\prime}&=-\Big(\frac{\partial g_k}{\partial s_j}\frac{\partial s_j}{\partial \Theta}\Big)\Big({\frac{\partial g_k}{\partial s_j}\dot s_j(\tau_k)}\Big)^{-1}
  \end{split}
  \end{equation}
At $\tau_k$ we have $\sigma_i(\tau_k)=\mu_{ij}P_{ij}(\tau_k)$. Therefore from \eqref{Xprime}:
\begin{equation}
\begin{split}
  &X_i^{\prime}(\tau_k^+)=X_i^{\prime}(\tau_k^-)+[\dot X_i(\tau_k^-)-\dot X_i(\tau_k^+)]{\tau_k}^{\prime}\\
  &=X_i^{\prime}(\tau_k^-)+\Big(0-\sigma_i(\tau_k)+\mu_{ij}P_{ij}(\tau_k)\Big){\tau_k}^{\prime}=X_i^{\prime}(\tau_k^-)
  \end{split}
  \label{xiplus}
\end{equation}
Having $X_i(t)>0$ in $[\tau_k,\tau_{k+1})$ we know $\dot X_i(t)=\sigma_i(t)-\mu_{ij}P_{ij}(t)$ therefor, we can get $\frac{d}{dt} X_i^{\prime}(t)$ from \eqref{dxdt} with $r$ and $l$ replaced by $i$ and $j$. For $X_r(t)$, $r \ne i$, if $X_r(\tau_k)>0$ and agent $l$ is connected to $r$ then $\dot X_r(\tau_k)=\sigma_r(\tau_k)-\mu_{rl}P_{rl}(\tau_k)$, therefor, we get $X^{\prime}_r(\tau_k^+)$ from \eqref{Xrprime0} while in $[\tau_k,\tau_{k+1})$ we have $\frac{d}{dt}X^{\prime}_r(t)$ from \eqref{dxdt}. If $X_r(\tau_k)=0$ or if no agent is connected to target $r$, $\dot X_r(\tau_k)=0$. Thus, $X^{\prime}_r(\tau_k^+)=X^{\prime}_r(\tau_k^-)$ and $ \frac{d}{dt}X^{\prime}_r(t)=0$.\\
For $Y_{r}(t)$, $r=1,\dots,M$ the dynamics of $Y_{r}(t)$ in \eqref{Ydot} are not affected by the event at $\tau_k$ hence, we can get $Y^{\prime}_{r}(\tau_k^+)$ and $\frac{d}{dt}Y^{\prime}_{r}(t)$ in $[\tau_k,\tau_{k+1})$ from \eqref{yrplus} and \eqref{dydt} respectively.\\
For $Z_{ij}(t)$ assuming agent $j$ is the one connected to target $i$, we have:
\begin{equation}
\begin{split}
  &Z^{\prime}_{ij}(\tau_k^+)=Z^{\prime}_{ij}(\tau_k^-)+\Big[\dot Z_{ij}(\tau_k^-)-\dot Z_{ij}(\tau_k^+)\Big]\tau^{\prime}_k\\
  &=Z^{\prime}_{ij}(\tau_k^-)+\Big[\tilde \mu_{ij}(\tau_k^-)-\tilde \mu_{ij}(\tau_k^+)\Big]P_{ij}(\tau_k)\tau^{\prime}_k=Z^{\prime}_{ij}(\tau_k^-)
  \end{split}
  \label{zijplus}
\end{equation}
In the above equation, $\tilde \mu_{ij}(\tau_k^+)=\mu_{ij}$ because $X_i(\tau_k^+)>0$. Also, $\mu_{ij}P_{ij}(\tau_k)=\sigma_i(\tau_k)$ and $\tilde \mu_{ij}(\tau_k^-)=\frac{\sigma_i(\tau_k)}{P_{ij}(\tau_k)}$ results in $\tilde\mu_{ij}(\tau_k^+)=\mu_{ij}$.
For $Z_{il}(t)$, $l\neq j$ , agent $l$ cannot be connected to target $i$ at $\tau_k$ so we have, $Z^{\prime}_{il}(\tau_k^+)=Z^{\prime}_{il}(\tau_k^-)$ and $\frac{d}{dt}Z^{\prime}_{il}(t)=0$ in $[\tau_{k}%
,\tau_{k+1})$.
For $Z_{rl}(t)$ ,$r\neq i$ and $l\neq j$ using the assumption that two events occur at the same time w.p. 0, the dynamics of $Z_{rl}(t)$ are not affected at $\tau_k$, hence
we get $\frac{d}{dt}Z^{\prime}_{rl}(t)$ from \eqref{dzdt} for $i$ and $j$ replaced by $r$ and $l$.\\

3. {\bf Event $\zeta_{ij}^{0}$}: This event causes a transition from $Z_{ij}(t)>0$ for $t<\tau_k$ to $Z_{ij}(t)=0$ for $t \ge\tau_k$. The switching function is $g_k(\Theta,{\bf X})=Z_{ij}(t)$ so $\frac{\partial g_k}{\partial Z_{ij}}=1$. From
\eqref{tauprime}:
\begin{equation}
\begin{split}
  &{\tau_k}^{\prime}=-\Big(\frac{\partial g_k}{\partial Z_{ij}}f_k(\tau^-_k)\Big)^{-1}\Big(\frac{\partial g_k}{\partial \Theta}+\frac{\partial g_k}{\partial Z_{ij}}{Z_{ij}^{\prime}(\tau^-_k)}\Big)\\
  &=-\frac{Z_{ij}^{\prime}(\tau^-_k)}{\tilde\mu_{ij}(\tau^-_k)P_{ij}(\tau^-_k)-\beta_{ij}P_{\!_{Bj}}(\tau^-_k)}=\frac{Z_{ij}^{\prime}(\tau^-_k)}{\beta_{ij}P_{\!_{Bj}}(\tau_k)}
\end{split}
\end{equation}
Since $Z_{ij}(t)$ is being emptied at $\tau_k$, by the assumption that agents can not be in range with the base and targets at the same time, we have $P_{ij}(\tau_k)=0$. Then from \eqref{Xprime}:
\begin{equation}
\begin{split}
 &Z^{\prime}_{ij}(\tau_k^+)=Z^{\prime}_{ij}(\tau_k^-)+\Big[-\beta_{ij}P_{\!_{Bj}}(\tau_k)-0\Big]{\tau_k}^{\prime}\\
 &=Z^{\prime}_{ij}(\tau_k^-)-\Big[\beta_{ij}P_{\!_{Bj}}(\tau_k)\Big]\frac{Z_{ij}^{\prime}(\tau^-_k)}{\beta_{ij}P_{\!_{Bj}}(\tau_k)}=0
 \end{split}
 \label{Zprime0}
\end{equation}
Since $\dot Z_{ij}(t)=0$ in $[\tau_k,\tau_{k+1})$:
\begin{equation}
  \frac{d}{dt}Z^{\prime}_{ij}(t)=\frac{\partial \dot Z_{ij}(t)}{\partial Z_{ij}(t)}Z^{\prime}_{ij}(t)+\frac{\partial{\dot Z_{ij}(t)}}{\partial \Theta}=0
  \label{Zprimet0}
\end{equation}
For $Z_{rl}(t)$, $r \ne i$ or $l \ne j$, the dynamics in \eqref{Zdot} are not affected at $\tau_k$, hence:
\begin{equation}
  Z^{\prime}_{rl}(\tau_k^+)=Z^{\prime}_{rl}(\tau_k^-)
  \label{Zrlplus}
\end{equation}
if $Z_{rl}(\tau_k)>0$, the value for $\frac{d}{dt}Z^{\prime}_{rl}(t)$ is calculated by \eqref{dzdt} with $r$ and $l$ replacing $i$ and $j$ respectively. If $Z_{rl}(\tau_k)=0$ then $ \frac{d}{dt}Z^{\prime}_{rl}(t)=0$. \\
For $Y_i(t)$ we have $\beta_{i}(\tau_k^+)=0$ since the agent has emptied its queue, hence:
\begin{equation}
\begin{split}
  &Y^{\prime}_{i}(\tau_k^+)=Y^{\prime}_{i}(\tau_k^-)+\Big[\dot Y_{i}(\tau_k^-)-\dot Y_{i}(\tau_k^+)\Big]\tau^{\prime}_k\\
  &=Y^{\prime}_{i}(\tau_k^-)+[\beta_{ij}P_{\!_{Bj}}(\tau_k)-0]\frac{Z_{ij}^{\prime}(\tau^-_k)}{\beta_{ij}P_{\!_{Bj}}(\tau_k)}\\
  &=Y^{\prime}_{i}(\tau_k^-)+Z_{ij}^{\prime}(\tau^-_k)
\end{split}
\label{Yprime}
\end{equation}
In $[\tau_k,\tau_{k+1})$ we can get $\frac{d}{dt}Y^{\prime}_{i}(t)=0$.
For $Y_{r}(t)$, $r\neq i$ the dynamics of $Y_{r}(t)$ in \eqref{Ydot} are not affected by the event at $\tau_k$ hence, $Y^{\prime}_{r}(\tau_k^+)$ and $\frac{d}{dt}Y^{\prime}_{r}(t)$ in $[\tau_k,\tau_{k+1})$ are calculated from \eqref{yrplus} and \eqref{dydt} respectively.
The dynamics of $X_r(t)$, $r=1,\dots,M$ is are not affected at $\tau_k$ since the event at $\tau_k$ is happening at the base. We have $X_r^{\prime}(\tau_k^+)=X_r^{\prime}(\tau_k^-)$.
If $X_r(\tau_k)>0$ then we have $\frac{d}{dt}X^{\prime}_r(t)$ from \eqref{dxdt} and if $X_r(\tau_k)=0$ then $\frac{d}{dt}X^{\prime}_r(t)=0$ in $[\tau_k,\tau_{k+1})$.\\

4. {\bf Event $\delta_{ij}^{+}$}: This event causes a transition from $D^{+}_{ij}(t)=0$ for $t\le\tau_k$ to $D^{+}_{ij}(t)>0$ for to $t>\tau_k$. It is the moment that agent $j$ leaves target $i$'s range. The switching function is $g_k(\Theta, {\bf  X})=D_{ij}(t)-r_{ij}$ , from \eqref{tauprime}:
\begin{equation}
 {\tau_k}^{\prime}=-
\frac{\partial D_{ij}}{\partial s_j}\frac{\partial s_j}{\partial \Theta}\Big(\frac{\partial D_{ij}}{\partial s_j}\dot s_j(\tau_k)\Big)^{-1}
\label{tauprime4}
\end{equation}
If agent $j$ was connected to target $i$ at $\tau_k$ then by leaving the target, it is possible that another agent $l$ which is within range with target $i$ connects to that target. This means $
  \dot X_i(\tau_k^+)=\sigma_i(\tau_k)-\mu_{il}P_{il}(\tau_k)$ and  $\dot X_i(\tau_k^-)=\sigma_i(\tau_k)-\mu_{ij}P_{ij}(\tau_k)$, with $P_{ij}(\tau_k)=0$, from \eqref{Xprime} we have
\begin{equation}
  X_i^{\prime}(\tau_k^+)= X_i^{\prime}(\tau_k^-)-\mu_{il}P_{il}(\tau_k)\tau_k^{\prime}
  \label{Xplusjump}
\end{equation}
If $X_i(\tau_k)>0$, $\frac{d}{dt}X^{\prime}_i(t)$ in $[\tau_k,\tau_{k+1})$ is as in \eqref{dxdt} with $r$ replaced by $i$ and if $X_i(\tau_k)=0$ then $\frac{d}{dt}X^{\prime}_i(t)=0$. On the other hand, if agent $j$ was not connected to target $i$ at $\tau_k$, we know that some $l\neq j$ is already connected to target $i$. This means agent $j$ leaving target $i$ cannot affect the dynamics of $X_i(t)$ so we have $X_i^{\prime}(\tau_k^+)=X_i^{\prime}(\tau_k^-)$ and $\frac{d}{dt}X_{i}^{\prime}(t)$ is calculated from \eqref{dxdt} with $r$ replaced by $i$.\\
For $X_r(t)$, $r \ne i$ the dynamics in \eqref{Xdot} are not affected by the event at $\tau_k$ hence, we get $X_r^{\prime}(\tau_k^+)$ from \eqref{Xrprime0}. If $X_r(\tau_k)>0$ the time derivative $\frac{d}{dt}X^{\prime}_r(t)$ in $[\tau_k,\tau_{k+1})$ can be calculated from \eqref{dxdt} and if $X_r(\tau_k)=0$ then $\frac{d}{dt}X^{\prime}_r(t)=0$.\\
For $Y_{r}(t)$, $r=1,\ldots,,M$, the dynamics in \eqref{Ydot} are not also affected by the event at $\tau_k$ hence, we get $Y_r(\tau_k^+)$ from \eqref{yrplus} and in $[\tau_k,\tau_{k+1})$ the  $\frac{d}{dt}Y^{\prime}_{r}(t)$ is calculated from \eqref{dydt}.\\
For $Z_{ij}(t)$, the dynamics in \eqref{Zdot} are not affect at $\tau_k$, regardless of the fact that agent $j$ is connected to target $i$ or not. We have $\dot Z_{ij}(\tau_k^-)=\tilde\mu_{ij}(\tau_k)P_{ij}(\tau_k)$ with $P_{ij}(\tau_k)=0$ and $\dot Z_{ij}(\tau_k^+)=0$, hence from \eqref{Xprime}:
\begin{equation}
\begin{split}
  &Z^{\prime}_{ij}(\tau_k^+)=Z^{\prime}_{ij}(\tau_k^-)+\Big[\dot Z_{ij}(\tau_k^-)-\dot Z_{ij}(\tau_k^+)\Big]\tau^{\prime}_k
  \\&=Z^{\prime}_{ij}(\tau_k^-)+\tilde\mu_{ij}(\tau_k)P_{ij}(\tau_k)\tau^{\prime}_k=Z^{\prime}_{ij}(\tau_k^-)
  \end{split}
\end{equation}
and in $[\tau_k,\tau_{k+1})$ , we have $\frac{d}{dt}Z^{\prime}_{ij}(t)=0$ using \eqref{dzdt} knowing $P_{ij}(\tau_k)=P_{\!_{Bj}}(\tau_k)=0$. For $Z_{rl}(t)$, $r \neq i$ or $l \neq j$, the dynamics of $Z_{rl}(t)$ are not affected at $\tau_k$ hence \eqref{Zrlplus} holds and in $[\tau_k,\tau_{k+1})$ again we can use \eqref{dzdt} with $i$ and $j$ replaced by $r$ and $l$.\\

5. {\bf Event $\delta_{ij}^{0}$}: This event causes a transition from $D^{+}_{ij}(t)>0$ for $t<\tau_k$ to $D^{+}_{ij}(t)=0$ for to $t\ge\tau_k$. The event is the moment that agent $j$ enters target $i$'s range. The switching function is $g_k(\Theta,{\bf X})=D_{ij}(t)-r_{ij}$. From \eqref{tauprime} we can get $ {\tau_k}^{\prime}$ from \eqref{tauprime4}. If no other agent is already connected to target $i$, agent $j$ connects to it. Otherwise, if another agent is already connected to target $i$, no connection is established. For $X_i(t)$, the dynamics in \eqref{Xdot} are not affected in both cases, hence, \eqref{xiplus} holds. If $X_i(t)>0$ in $[\tau_k,\tau_{k+1})$ we calculate $\frac{d}{dt}X^{\prime}_i(t)$ using \eqref{dxdt} with $l$ being the appropriate connected agent to target $i$. If $X_i(\tau_k^-)=0$, $\frac{d}{dt}X^{\prime}_i(t)=0$.
For $X_r(t)$, $r \ne i$ the dynamics in \eqref{Xdot} are not affected by the event at $\tau_k$. Hence, we get $ X_r^{\prime}(\tau_k^+)$ from \eqref{Xrprime0}. If $X_r(\tau_k)>0$  we calculate $\frac{d}{dt}X^{\prime}_r(t)$ from \eqref{dxdt} with $i$ replaced by $r$ and if $X_r(\tau_k)=0$ then $\frac{d}{dt}X^{\prime}_r(t)=0$.\\
For $Y_{r}(t)$, $r=1,\dots,M$ again the dynamics in \eqref{Ydot} are not affected at $tau_k$ so both \eqref{yrplus} and \eqref{dydt} hold.\\
For $Z_{ij}(t)$, with agent $j$ being connected or not to target $i$ at $\tau_k$ the dynamics of $Z_{ij}(t)$ are unaffected at $\tau_k$, hence \eqref{Zrlplus} holds for $i$ and $j$ and in $[\tau_k,\tau_{k+1})$ the $\frac{d}{dt}Z^{\prime}_{ij}(t)$ is calculated through \eqref{dzdt}. For $Z_{rl}(t)$, $r \neq i$ or $l \neq j$  the dynamics are unaffected \eqref{Zrlplus} holds again. In $[\tau_k,\tau_{k+1})$, $\frac{d}{dt}Z^{\prime}_{rl}(t)$ is given through \eqref{dzdt} with $i$ and $j$ replaced by $r$ and $l$.\\

6. {\bf Event $\Delta_{j}^{+}$}: This event causes a transition from $D^{+}_{\!{Bj}}(t)=0$ for $t\le\tau_k$ to $D^{+}_{\!{Bj}}(t)\ge 0$ for $t> \tau_k$. The switching function is $g_k(\Theta,{\bf X})=D_{\!_{Bj}}(t)-r_{\!_{Bj}}$.
\begin{equation}
 {\tau_k}^{\prime}=-\frac{\partial D{\!_{Bj}}}{\partial s_j}\frac{\partial s_j}{\partial \Theta}\Big(\frac{\partial D_{\!_{Bj}}}{\partial s_j}\dot s_j(\tau_k)\Big)^{-1}
\label{tauprime6}
\end{equation}
Similar to the previous event, the dynamics of $X_i(t)$ are unaffected at $\tau_k$ hence, we have $X_i^{\prime}(\tau_k^+)$ calculated from \eqref{xiplus}.
If $X_i(t)>0$ in $[\tau_k,\tau_{k+1})$ we calculate $\frac{d}{dt}X^{\prime}_i(t)$ through \eqref{dxdt} and if $X_i(\tau_k^-)=0$, $\frac{d}{dt}X^{\prime}_i(t)=0$.\\
For $Y_{r}(t)$, $r=1,\ldots,,M$, the dynamics of $Y_r(t)$ in \eqref{Ydot} are not affected at $\tau_k$, hence, we get $Y_r(\tau_k^+)$ from \eqref{yrplus}
and in $[\tau_k,\tau_{k+1})$, $ \frac{d}{dt}Y^{\prime}_{r}(t)$ is calculated from \eqref{dydt}.\\
For $Z_{ij}(t)$, Using the fact that agent $j$ can only be connected to one target or the base, we have $\dot Z_{ij}(\tau_k^-)=\beta_{ij}(\tau_k)P_{\!_{Bj}}(\tau_k)$ with $P_{\!_{Bj}}(\tau_k)=0$ and $\dot Z_{ij}(\tau_k^+)=0$, hence \eqref{Zrlplus} holds with $i$ and $j$ replacing $r$ and $l$.
In $[\tau_k,\tau_{k+1})$ from \eqref{dXdtprime}:
\begin{equation}
\begin{split}
  &\frac{d}{dt}Z^{\prime}_{ij}(t)=\frac{\partial \dot Z_{ij}(t)}{\partial Z_{ij}(t)}Z^{\prime}_{ij}(t)+\frac{\partial{\dot Z_{ij}(t)}}{\partial \Theta}\\&=\frac{\partial{\dot Z_{ij}(t)}}{\partial \Theta}=-\beta_{ij} P^{\prime}_{\!_{Bj}}(t)
  \end{split}
  \label{ZprimetB}
\end{equation}
 As for $Z_{rl}(t)$, $r \neq i$ or $l \neq j$ the dynamics are unaffected so  \eqref{Zrlplus} holds. In $[\tau_k,\tau_{k+1})$ we can calculate $\frac{d}{dt}Z^{\prime}_{rl}(t)$ through \eqref{dzdt} with $j$ replacing $l$.\\

7. {\bf Event $\Delta_{j}^{0}$}: This event causes a transition from $D^{+}_{\!{Bj}}(t)>0$ for $t<\tau_k$ to $D^{+}_{\!{Bj}}(t)=0$ for $t\ge \tau_k$. The switching function is $g_k(\Theta, {\bf X})=D_{\!_{Bj}}(t)-r_{\!_{Bj}}$. Using \eqref{tauprime} we can get
$ {\tau_k}^{\prime}$ from \eqref{tauprime6}. Similar with the previous event we have $X_i^{\prime}(\tau_k^+)$ from \eqref{xiplus}. If $X_i(t)>0$ we can get $\frac{d}{dt}X^{\prime}_i(t)$ from \eqref{dxdt} and if $X_i(\tau_k^-)=0$ then $\frac{d}{dt}X^{\prime}_i(t)=0$.\\
For $Y_{r}(t)$, $r=1,\ldots,,M$, we again follow the previous event analysis so \eqref{yrplus} and \eqref{dydt} hold.\\
For $Z_{ij}(t)$, the analysis is similar to event $\Delta_j^+$ so we can calculate $Z_{ij}^{\prime}(\tau_k^+)$ and $\frac{d}{dt}Z^{\prime}_{ij}(t)$ in $[\tau_k,\tau_{k+1})$ from \eqref{zijplus} and \eqref{dzdt} respectively.
Also for $Z_{rl}(t)$, $r \neq i$ or $l \neq j$, \eqref{Zrlplus} holds with same reasoning as previous event. In $[\tau_k,\tau_{k+1})$ we calculate $\frac{d}{dt}Z^{\prime}_{rl}(t)$ from \eqref{dzdt}.
\end{appendices}

\bibliographystyle{hieeetr}
\bibliography{C:/Users/Yas/Dropbox/Academic/Research/Fall2014/DH/DH_IPA_Paper/Ref_DH}

\end{document}